\documentclass[preprint,12pt,authoryear]{elsarticle}
\usepackage[english]{babel}
\usepackage[utf8]{inputenc}
\usepackage[T1]{fontenc}
\usepackage{graphicx, float} 
\usepackage{amssymb}
\usepackage{amsmath}
\usepackage{graphicx, color}
\usepackage{natbib}
\usepackage{subcaption} 
\usepackage{multirow}
\usepackage{url} % not crucial 
\usepackage{makecell}

\begin{document}

\begin{frontmatter}
	
	\title{Value-Aware Product Recommendation by Customer Segmentation using a suitable High-Dimensional Similarity Measure} 
	
	\author[label5]{María Florencia Acosta} 
	\ead{mfacosta@santafe-conicet.gov.ar}
	\author[label1,label3]{Rodrigo García Arancibia} 
	\ead{rgarcia@fce.unl.edu.ar}
	\author[label1,label2]{Pamela Llop\corref{cor1}}
	\author[label1,label2]{Mariel Lovatto} 
	\ead{mlovatto@fiq.unl.edu.ar}
	\author[label1,label4]{Lucas Mansilla} 
	\ead{lmansilla@sinc.unl.edu.ar}
	
	\cortext[cor1]{Corresponding author. Email: pllop@santafe-conicet.gov.ar}

	\affiliation[label1]{organization={Consejo Nacional de Invesitgaciones Científicas y Técnicas (CONICET)},
		city={Santa Fe},
		postcode={3000}, 
		state={Santa Fe},
		country={Argentina}}     
	\affiliation[label2]{organization={Facultad de Ingeniería Química, Universidad Nacional del Litoral},
		city={Santa Fe},
		postcode={3000}, 
		state={Santa Fe},
		country={Argentina}}    
	\affiliation[label3]{organization={Instituto de Economía Aplicada Litoral, Universidad Nacional del Litoral},
		city={Santa Fe},
		postcode={3000}, 
		state={Santa Fe},
		country={Argentina}}
	\affiliation[label4]{organization={Instituto de Investigación en Señales, Sistemas e Inteligencia Computacional, sinc($i$), Universidad Nacional del Litoral},
		city={Santa Fe},
		postcode={3000},
		state={Santa Fe},
		country={Argentina}}
	\affiliation[label5]{organization={Facultad de Ingeniería y Ciencias Hídricas, Universidad Nacional del Litoral},
		city={Santa Fe},
		postcode={3000},
		state={Santa Fe},
		country={Argentina}}
	
		%% Keywords

%% Abstract
\begin{abstract}
This paper presents a novel value-aware approach to product recommendation that simultaneously addresses the high dimensionality and sparsity of user–item data while explicitly incorporating the contribution of each product and user to overall sales revenue. The proposed framework encodes revenue contributions in the user–item matrix and computes customer similarity directly on this basis using suitable distance measures. This enables the segmentation of users according to the revenue-based similarity of their purchase baskets and supports recommendations aligned with profitability objectives. We compare conventional similarity metrics with a new novel alternative tailored to high-dimensional contexts and propose three recommendation strategies: based on revenue share, product popularity, and expected profit generation. The effectiveness of the proposed method is validated through simulation experiments and a real-world application using the UCI Online Retail dataset.

\end{abstract}


% \begin{highlights}
% \end{highlights}

%% Keywords
%\begin{keyword}



%\end{keyword}

\end{frontmatter}

\section{Introduction}
In recent decades, the rapid proliferation of automated recommendation systems (RSs) has fundamentally transformed the way users discover information and products. From tailoring film and music suggestions to personalizing news consumption and shopping experiences, the influence of RSs has become pervasive, raising new challenges regarding personalization, user autonomy, and information diversity \citet{fayyaz2020recommendation,alamdari2020systematic,alves2023review, ko2022survey}. Their utility has now expanded beyond commercial applications, reaching domains as diverse as the recommendation of public policies for sustainable development goals \citet{felfernig2023recommender}.

Recommendation techniques are commonly categorized into four main types: content-based, collaborative filtering (CF), knowledge-based, and hybrid approaches, with their applicability depending on the specific context \citet{xia2024contemporary}. Among these, CF remains one of the most prominent and successful paradigms \citet{su2009survey, yildiz2023hyper}. It operates by generating recommendations based solely on users' historical behavior, such as past purchases or item ratings \citet{debiasio2024}. A particularly powerful variant of CF exploits user similarity derived from behavioral data, like purchase histories or basket analysis 
\citet{alves2023review, CHRISTY20211251}. By identifying clusters of customers/users with analogous preferences, these systems have achieved high popularity due to their simplicity, efficiency, and scalability in accurately  predicting items that appeal to similar individuals or a user's nearest neighbors \citet{aggarwal2016recommender,nikolakopoulos2021trust}. 

A crucial step in RSs, particularly in segmentation-based CF, is the identification of relevant customer data. User behavior patterns are derived from this, which can be collected either through explicit or implicit means. Explicit feedback is directly provided by the user, such as likes and ratings on smart devices. On the other hand, implicit information data is inferred from user actions, such as the analysis of shopping basket data from purchases in both physical and online stores. The nature and scope of the available data, often sourced from multi-domain environments, ultimately guide the selection of appropriate techniques and algorithms, and can even motivate the need for novel methodological developments to enhance the personalization of recommendations \citet{ko2022survey}.

For clustering-based techniques, selecting an appropriate similarity measure is essential. This choice should be driven by the nature of the data in the user-interaction matrix, which may be continuous, interval-based, ordinal categorical, or binary \citet{aggarwal2016recommender}. Some distance metrics are more appropriate or/and used for certain data types, as the popular Pearson correlation, euclidean or cosine similarity measures for continuous and rating data \citet{hossain2017customer,8371003} or Jaccard distance for binary data \citet{bag2019efficient,verma2020comparative,bae2025ranking}. Consequently, the choice of an appropriate similarity measure directly affects clustering accuracy and, ultimately, the overall performance of RSs \citet{xiaojun2017improved,fkih2022similarity}. Nevertheless, the practical effectiveness of these similarity measures is often critically influenced by the density and dimensionality of the data. In many real-world scenarios, RSs must operate on matrices with severe sparsity and high dimensionality.

Despite the increasing availability of data and the significant advances in RSs, {\it  data scarcity} remains one of the most persistent challenges in the field of RSs, as is noted by \cite{chen2025data}. This problem is usually reflected in a highly sparse user-item interaction matrix, which severely limits the effectiveness of recommendation models. Sparsity is more frequent in high-dimensional context, that is, with a large number of items relative to users. This occurs because user behavior tends to be concentrated on a few popular items, while the vast majority receive little to no interaction. Moreover, over time, the total catalog of items is typically much larger than the number a single user interacts with,  naturally resulting in a sparse matrix structure. Data scarcity and sparsity in RSs have been primarily addressed through alternative learning approaches, such as data augmentation \citet{bansal2022systematic}, self-supervised learning \citet{yu2023self}, transfer learning \citet{zhao2013active}, broad learning \citet{zhu2017broad}, and knowledge graphs \citet{shao2021survey}; for an extensive review on this topic, see for instance \citet{chen2025data} .

%hablar de las soluciones, que no tienen enn cuenta cambiar la metrica sino los metodos. Nombrar algunso métodos de ese paper, por ejemplo. 

In contrast to these more complex methodologies, in this paper we introduce a simple approach that adjusts conventional distance metrics, such as Euclidean or cosine, to avoid the concentration of pairwise distances and then mitigating the resulting degradation of neighborhood structure. The new metric is the so-called Mean of Absolute Differences of pairwise Distances (MADD) \citet{sarkar2020}. Unlike the mentioned metrics, MADD exploits the distance concentration phenomenon, resulting in superior clustering performance in high-dimensional spaces. In the context of RSs, our goal is to enhance the identification of users with similar preferences, thereby improving the foundation for product recommendations without altering the underlying clustering framework, and for this objective, MADD shows to have great properties.

Beyond the role of data type in selecting similarity metrics for clustering, a critical consideration for firms is integrating value-aware dimensions into the user-interaction matrix \citet{chen2008developing, garcin2014offline, lu2014show, panniello2016impact, abdollahpouri2020multistakeholder, debiasio2023}. For instance, by incorporating data such as customer revenue contribution, the clustering process can group users not only by preference similarity but also by their comparable impact on business outcomes like sales revenue. This strategic approach, termed value-aware clustering, shifts the focus from mere preference alignment to customer value alignment. The goal is a balanced outcome where firms increase profitability while customers receive more valuable and relevant recommendations \citet{cai2019trustworthy, nemati2020devising, kompan2021exploring, concha2023multi}. When this value-centric segmentation is coupled with a top-$L$ value maximization strategy, which prioritizes items that maximize expected revenue over predicted rating, the result is a more powerful system directly optimized for enhancing sales and firm profitability \citet{debiasio2024, de2024economic}.

Taking these considerations into account, this paper proposes a novel product RS based on clustering consumer baskets. Each basket is represented by a user's share of spending on each product relative to the total expenditure across all users and products. This user-interaction matrix is designed to capture customer budget allocation and the contribution of each product to sales revenue. A key advantage of this representation is its inherent stability under inflationary contexts, as it relies on expenditure shares rather than absolute prices, eliminating the need for a price deflator index. To address the challenges of sparse and high-dimensional data, we employ the MADD similarity measure, which is well-suited for this context. Finally, the recommendation step is formulated as a top-$L$ value maximization problem, incorporating alternative ranking criteria such as product popularity, revenue share, and expected profit. We evaluate the performance of our proposed system through experiments on both simulated data, under various alternative scenarios, and real-world data from the Online Retail II dataset. The evaluation compares our method against standard similarity metrics commonly used in segmentation-based RSs, focusing on both clustering effectiveness and value-aware recommendation performance. Our code is publicly available at \url{https://anonymous.4open.science/value-aware-recsys}.

\section{Methodology} \label{methodolog}
As mentioned earlier, this paper focuses on the collaborative filtering algorithm for product recommendations, which can be divided into three stages \cite{beregovskaya2021}:
\begin{itemize}
\item {\bf Building the user model}: Represented by a user-item interaction matrix, where each cell captures the relationship between a specific user and item.
\item {\bf Identifying the closest set of neighbors}: Involves applying clustering techniques and similarity measures to find users with similar interaction patterns.
\item {\bf Generating recommendations}: Based on the identified neighbors, a criterion is defined to produce a list of recommended items.
\end{itemize}

These three steps are explained in detail in the following subsections.

\subsection{Building the user model}
Consider a firm offering a set of $p$ distinct products $\mathcal{J} = \{1, \dots, p\}$ to a population of $n$ customers $\mathcal{U} = \{1, \dots, n\}$. For each customer-product pair $(u, j) \in \mathcal{U} \times \mathcal{J}$, the firm observes an expenditure $m_{uj} \in [0, E]$, where $E = \max_{u,j} m_{uj}$ represents the maximum expenditure observed during a fixed time period. This expenditure decomposes as  $m_{uj} = q_{uj} \cdot p_{uj}$, where $q_{uj}$ denotes the quantity sold of the product $j$ to the user $u$ and $p_{uj}$ its unit price. The price $p_{uj}$ may vary across customers due to personalized pricing or temporal fluctuations. 

When implementing recommendation systems, the nominal expenditure matrix $\mathbf{M}$ with entries $[\mathbf{M}]_{uj} = m_{uj}$ presents challenges in high-inflation environments. %Temporal price-level changes cause identical consumption baskets to appear dissimilar in nominal terms, while asymmetric price adjustments across products further distort preference signals. 
Temporal changes in price levels cause identical consumption baskets to appear different in nominal terms, while asymmetric price adjustments across products further distort the underlying preference signals.
Deflating expenditures to real terms via $\widetilde{m}_{uj}=m_{uj}/P_t$, where  $P_t$ is a period-specific price index, offers a theoretical solution but introduces practical complications: it requires timely access to granular price indices, imposes maintenance overhead for index updates, and assumes homogeneous inflation across product categories, conditions rarely satisfied in practice. %This tension between theoretical rigor and operational feasibility underscores the need for robust recommendation methods that mitigate inflationary distortions while maintaining implementability.
This tension between theoretical rigor and practical feasibility highlights the need for robust recommendation methods that effectively mitigate inflationary distortions while remaining implementable. 

Alternatively, one might consider using the \emph{real} quantity \( q_{uj} \), defining the interaction matrix \( \mathbf{Q} \) with \( [\mathbf{Q}]_{uj} = q_{uj} \). However, this approach has limitations because quantities are often measured in incompatible units (e.g., liters versus kilograms), making meaningful comparisons across products difficult. 

Another  common choice in recommendation via collaborative filtering,  which is independent of monetary values or measurement units, is the \textbf{binary user-item interaction matrix} \( \mathbf{D} \), where \( [\mathbf{D}]_{uj} =d_{uj} = 1 \) if \( m_{uj} > 0 \) and \( 0 \) otherwise. While this matrix effectively captures purchase incidence, it ignores valuable information about expenditure levels that could help identify distinct customer segments (e.g., bulk buyers of low-priced items vs. niche buyers of premium products).

To overcome these limitations, we propose using the {\bf sales revenue share} of product $j$ purchased by customer $u$, defined as:
\begin{equation}
    s_{uj} = \frac{m_{uj}}{\sum_{u=1}^n \sum_{j=1}^p m_{uj}}.
\end{equation}
This definition produces the \textbf{share user-item interaction matrix} $\mathbf{S}$, where $[\mathbf{S}]_{uj} = s_{uj}$. By normalizing expenditures in this way, the matrix preserves meaningful distinctions across both customers and products. 

The revenue share approach captures three key dimensions for enhanced recommendations: (i) \textit{client importance} through their relative contribution to total sales ($\sum_j s_{uj}$), (ii) \textit{spending patterns} via the distribution across products ($s_{u1},\ldots,s_{up}$), and (iii) \textit{product profitability} through items' revenue contributions ($\sum_u s_{uj}$). Here, each $s_{uj}$ represents the proportion of total revenue attributable to customer $u$'s purchases of product $j$. This formulation naturally embeds profit-awareness by weighting recommendations according to actual sales impact, while remaining robust to inflationary distortions that affect nominal expenditures. This approach aligns with profit-aware recommendation frameworks \cite{jannach2017price,bae2025ranking} while addressing inflation-related challenges.

On the other hand, consistent with many real-world applications, we assume a high-dimensional and sparse setting. This imposes a specific structure on the  $n\times p$ matrix $\mathbf{S}$, where typically $p \gg n$, meaning the number of items far exceeds the number of customers available for training a recommendation system. As a result, the matrix contains a large proportion of zero entries. Consequently, when comparing customer baskets (rows of $\mathbf{S}$) of different customers, these vectors exhibit high sparsity, necessitating specialized similarity metrics to accurately assess their (dis)similarity in a user-based collaborative filtering (CF) framework. To address this challenge, we propose a tailored similarity measure and evaluate its performance against conventional metrics used in this context.

For the customer $u\in \mathcal{U}$, we will use the notation $\mathbf{v}_u$ to identify either, the $p$-dimensional row $\mathbf{d}_u$ of the binary user-item interaction matrix $[\mathbf{D}]_{uj}$ or $\mathbf{s}_u$ the $p$-dimensional row of the share user-item interaction $[\mathbf{S}]_{uj}$. This is,
\begin{equation}\label{values}
  \mathbf{v}_u \doteq \begin{cases}
    \mathbf{d}_u & \text{ if } \, [\mathbf{D}]_{uj},\\
    \mathbf{s}_u & \text{ if } \,  [\mathbf{S}]_{uj}.
\end{cases}  
\end{equation}

\subsection{Identifying the closest set of neighbors}
%\subsection{Segmenting customers}

To perform recommendation, we first segment the set of customers into $K$ clusters $C_k$ for $k=1,\ldots,K$ via $k$-medoids algorithm. In this direction, we employ the \textbf{k-medoids algorithm} \citet{kaufman1987}. This clustering technique, similar to k-means, partitions the dataset into $k$ groups of similar users, this is, users whose baskets are similar. Unlike $k$-means, in the k-medoids algorithm each cluster is represented by a data point called a \textbf{medoid}, which is the most centrally located object within the cluster. A key advantage of this approach in the present context is that the medoid represents an actual consumption basket purchased by a specific customer, rather than an artificial mean basket that may not correspond to any real consumer's behavior. By using the $k$-medoids algorithm, all distance measures are computed between observed consumption baskets, ensuring that the resulting clusters are grounded in real purchasing patterns. This property is particularly valuable in market segmentation, where actionable insights require profiles based on genuine consumer behavior. In addition, it can work with any distance or dissimilarity measure, which makes $k$-medoids more robust to outliers and better suited for non-Euclidean spaces. We will present the different similarities measures used in this work in next section. 

One of the most commonly used algorithms for performing $k$-medoids clustering is \textbf{PAM} (Partitioning Around Medoids) \citet{kaufman1987, kaufman1990}. It consists of finding representative objects, the medoids, among the observations in the dataset. Clusters are formed by assigning each observation to the nearest medoids. Then, the algorithm performs a swap step to improve the clustering quality by exchanging selected medoids with non-selected objects. This process continues until the objective function, which measures the dissimilarities of objects to their nearest medoids, can no longer be decreased. The goal is to find $k$ representative objects that minimize the sum of dissimilarities to their nearest medoids.

For $\mathbf{v}_u$, the value defined in \eqref{values} for customer $u \in \mathcal{U}$, we assign the cluster $C_k$ for $k=1,\ldots,K$ via minimal dissimilarity to the corresponding medoids $\mathbf{v}^m_k$. Specifically, 
\begin{equation}
C_k = \underset{k \in \{1,\ldots,K\}}{\arg\min} \, \delta(\mathbf{v}_u, \mathbf{v}^m_k)
\end{equation}
where $\delta(\cdot, \cdot)$ stands for the four distances introduced in \eqref{euclidean} to \eqref{madd}. 

To determine the appropriate number of clusters $k$, we use the \textbf{Silhouette coefficient} ($\mathcal{S}$), which evaluates how well each object lies within its cluster. A higher average silhouette score suggests a more appropriate clustering structure. More precisely, for $u=1,\ldots, n$, we define the coefficient $\mathcal{S}$ as,
\[
\mathcal{S}(u) = \frac{B(u) - A(u)}{\max\{A(u),\, B(u)\}},
\]
where $A(u)$ is the average distance from customer $u$ to all other users in the same cluster, and $B(u)$ is the average distance from $u$ to the nearest cluster to which it does not belong.

As discussed before, $k$-medoids clustering requires defining a similarity measure between purchased baskets of customers $u$ and $v$ in  $\mathcal{U}$. These similarity measures rely on distances between the values $\mathbf{v}_u$ and $\mathbf{v}_v$ defined in \eqref{values} and are presented in the next section.

\subsubsection{Similarity measures}\label{similarity-measures}

In recommendation systems, the most commonly used similarity measures are based on the {Euclidean} and {Cosine} distances, with the latter being preferred for high-dimensional and sparse data due to its invariance to vector magnitude. However, as noted by \cite{peng2024} both measures suffer from the {\it distance concentration} a phenomenon where pairwise distances between data points converge to similar values as dimensionality grows, degrading the ability to distinguish meaningful clusters or neighborhoods. To mitigate this issue, specialized high-dimensional similarity measures have been proposed for clustering and related tasks \citet{sarkar2020, modarres2022}. In this work, we employ a similarity measure based on the {MADD} (Mean Absolute Difference of Distances) metric. Although MADD is not traditionally used in recommendation systems for customer segmentation, it exhibits advantageous properties for this application, demonstrating superior clustering performance and, consequently, enhanced recommendation quality in high-dimensional settings. We also evaluate measures based on the {Jaccard} distance, which is widely adopted in recommendation systems for binary or categorical user-item interaction matrices. A modern reference for Cosine and Jaccard metrics is the book \cite{tan2018}. 

In what follows, we present the four similarity measures compared in this work. Let $\mathbf{v}_u$ and $\mathbf{v}_v$ the values defined in \eqref{values} for customer $u$ and $v$ in  $\mathcal{U}$, respectively. For these two values, we define the following similarities measures:

\begin{itemize}
    \item The \textbf{Euclidean similarity} \citet{johnson2007}: based on the classical euclidean distance between vectors 
    \begin{equation}\;\label{euclidean}
    d(\mathbf{s}_u, \mathbf{s}_v) = \sqrt{\sum_{j=1}^p (s_{u,j} - s_{v,j})^2}.
    \end{equation}
    \item The \textbf{Cosine similarity} \citet{manning2008}: based on the cosine of the angle $\theta$ between the vectors $\mathbf{s}_u$ and $\mathbf{s}_v$. Is mathematical formula us computed from the inner product between them and it is given by 
    \begin{equation}\label{cosine}
     \text{cos}(\mathbf{s}_u, \mathbf{s}_v) = 1-\frac{\mathbf{s}_u \cdot \mathbf{s}_v}{||\mathbf{s}_u|| ||\mathbf{s}_v||},   
    \end{equation}    
    where $\cdot$ states for the inner product between the vectors and and $||\cdot||$ for the norm of them.
    \item The \textbf{Jaccard similarity} \citet{jaccard1901, tan2018}: it is defined as  
    \begin{equation}\label{jaccard}
     J(\mathbf{d}_u, \mathbf{d}_v) = 1-\frac{|\mathbf{d}_u \cap \mathbf{d}_v|}{|\mathbf{d}_u \cup \mathbf{d}_v|},  
    \end{equation}
    where $|\mathbf{d}_u \cap \mathbf{d}_v|$ indicates the number of elements that baskets of users $u$ and $v$ have in common and $|\mathbf{d}_u \cup \mathbf{d}_v|$ is the total number of distinct elements in both baskets. 
     \item The \textbf{MADD similarity} \citet{sarkar2020}: based on the \textit{Mean Absolute Difference of Distances} between $\mathbf{s}_u$ and $\mathbf{s}_v$, is given by 
    \begin{equation}\label{madd}
     \text{MADD}(\mathbf{s}_u, \mathbf{s}_v) = \frac{1}{n - 2} \sum_{\substack{l = 1 \\ l\ne u,\, l \ne v}}^{n} \left| d(\mathbf{s}_u, \mathbf{s}_l) - d(\mathbf{s}_v, \mathbf{s}_l) \right|,   
    \end{equation}    
    where $d$ states for the euclidean distance (Eq. \eqref{euclidean}) between $\mathbf{s}_u$ and $\mathbf{s}_v$, and $|\cdot|$ is the absolute value between both distances.
\end{itemize}
\subsubsection{Considerations about Computational Cost}
The choice of distance metric directly impacts the computational cost of constructing the similarity matrix for clustering. We analyze the complexity of the metrics considered in this study.

{The Euclidean distance requires $\mathcal{O}(p)$ operations for a single costumer pair, leading to a total cost of $\mathcal{O}(n^2p)$ for the full similarity matrix. The cosine distance also has a theoretical cost of $\mathcal{O}(n^2p)$, though practical implementations can achieve lower constants by pre-computing user vector norms ($||\mathbf{s}_u||$) to avoid redundant calculations. In contrast, the Jaccard distance offers a lower computational cost of $\mathcal{O}(n^2k)$, where $k \ll p$ represents the average number of items with which a user interacts. This makes the Jaccard similarity an efficient alternative in sparse data contexts. However, this efficiency comes at a cost: the Jaccard metric discards valuable information contained in the user interaction matrix, such as revenue shares or interaction frequencies.}

{Finally, the MADD distance incurs a significantly higher computational cost. For each user pair $(u, v)$, comparisons must be made against all other $(n-2)$ users, with each elementary distance calculation costing $\mathcal{O}(p)$. This results in a total complexity of $\mathcal{O}(n^3p)$. Consequently, MADD is computationally intensive and generally unsuitable for datasets with millions of users.
Despite its higher complexity, the $\mathcal{O}(n^3p)$ cost of MADD is not prohibitive in the context of our analysis. In many recommendation system settings, the number of items $p$ is considerably larger than the number of users $n$ ($p \gg n$). Furthermore, clustering is often performed on user segments of moderate size.}

{Therefore, the increased computational cost is offset by the significant benefits MADD provides, namely its robustness against the distance concentration phenomenon and its superior ability to preserve geometric structure in high-dimensional spaces. Thus, the use of MADD is justified in scenarios where clustering quality is prioritized over computational speed.}

\subsection{Generating recommendations} \label{recommendation-methods}

Once defined the cluster $C_k$ to which $u$ belongs, the final step is to use different ranking criteria to generate a top-$L$ recommendation list based either on expected interest and item \textit{popularity} (classical top-$L$ recommendation problem) or on economic values such as \textit{revenue shares} (referred to as the top-$L$ value maximization problem). To do it, let us first introduce some notation. 

For customer $u\in \mathcal{U}$ let us define its  \textbf{purchased basket} $\beta_u$ among all items $\mathcal{J}$ as,
\[
\beta_u \doteq \{j \in \mathcal{J} : j \text{ is purchased by user } u \in \mathcal{U}\},
\]
%Let $u\in \mathcal{U}$ the \textit{target} consumer with a current consumer basket $\mathbf{b}_u$ with the corresponding values $\mathbf{v}_u$ in the user-item interaction matrix (i.e. $\mathbf{v}_u=\mathbf{s}_u,\mathbf{d}_u$), and to which we want to recommend $L$ non-purchased products. The recommendation system first assigns $u$ to a cluster $C_k$ for $k=1,\ldots,K$ via minimal dissimilarity to the corresponding medoids $\mathbf{v}^m_k$.
%Let $u\in \mathcal{U}$ the \textit{target} consumer to which we want to recommend products. This user has a current purchased basket $\mathbf{b}_u$ with the corresponding values $\mathbf{v}_u$ in the user-item interaction matrix 
and let $\mathcal{J}_k \subseteq \mathcal{J}$ be the set of products purchased by all users in cluster $C_k$, that is
\begin{equation*}
\mathcal{J}_k = \bigcup_{u \in C_k} \beta_u.
\end{equation*}
% \begin{equation*}
% \mathcal{J}_k = \{ j \in \mathcal{J} \,|\, \exists u \in C_k \text{ such that } j \in \mathbf{b}_u \}
% \end{equation*}
%where $\mathbf{b}_u$ denotes the consumption basket of user $u$. 
For each product $j \in \mathcal{J}_k$, we define its frequency of purchase as
\begin{equation*}
% f_{j,k} = |\{ u \in C_k : j \in \mathbf{b}_u \}| \doteq \sum_{u\in C_k} d_{uj},
f_{j,k} \doteq \sum_{u\in C_k} d_{uj},
\end{equation*}
%$|\cdot|$ stands for the cardinal of the set and 
where $d_{uj}$ is the element $(u,j)$ of the binary user-item interaction matrix $\mathbf{D}$. 
%The first step of the recommendation system consists in allocate the costumer $u^t$ in a cluster $C_k$ for $k=1,\ldots,K$ via the minimal distance with the corresponding medoids. That is, $d(\mathbf{b}_u,\mathbf{b}^m_k)$, where $\mathbf{b}^m_k$
% is the medoids of cluster $k$. Given the assigned cluster $C_k$, let  $\mathcal{J}_k \subseteq \mathcal{J}$ be the set of products purchased by users in cluster $C_k$:
%\begin{equation}
%\mathcal{J}_k = \{ j \in \mathcal{J} \,|\, \exists u \in C_k \text{ such that } j \in \mathbf{b}_u \}
%\end{equation}
%where $\mathbf{b}_u$ denotes the consumption basket of user $u$. For each product $j \in \mathcal{J}_k$, we define the purchase frequency:
%\begin{equation*}
%f_{j,k} = \left| \{ u \in C_k : j \in \mathbf{b}_u \} \right|
%\end{equation*}

Then, we consider the following three alternative scoring functions $V$ for the top-$L$ value maximization problem:
\begin{enumerate}
    \item \textbf{Popularity-based score:} Measures the popularity of the product $j$ within cluster $C_k$ by computing its relative frequency in the cluster,
    \begin{equation}
    V_{\texttt{pop}}(j, C_k) = \frac{f_{j,k}}{|C_k|}
    \end{equation}
    where $|C_k|$ denotes the number of users in cluster $C_k$ (i.e., the cardinality of set $C_k$).
    
    \item \textbf{Revenue-based score:} Captures the revenue contribution of item $j$ in the cluster $C_k$ through its revenue share
    \begin{equation}
    V_{\texttt{rev}}(j, C_k) = s^k_j
    \end{equation}
    where $s^k_j$ represents the revenue share of product $j \in \mathcal{J}_k$. This is, $s^k_j=[\mathbf{S}]_{u,j}$ for some $u\in C_k$.
    
    \item \textbf{Expected Profit score:} Quantifies expected profit generation by combining popularity and revenue contribution
    \begin{equation}
    V_{\texttt{exppro}}(j, C_k) = \frac{f_{j,k}}{|C_k|} \cdot s^k_j.
    \end{equation}
    This score reflects the expected revenue (or profit) generated by recommending the product $j$ to a random user in cluster $C_k$.
\end{enumerate}
Finally, for a chosen scoring function $V_{\ell}(j, C_k)$ with $\ell= \{\texttt{pop,rev,exppro}\}$, the top-$L$ recommendation problem is formulated as

\begin{equation}
\begin{aligned}
& \underset{\mathcal{L} \subseteq \mathcal{J}_k}{\text{maximize}} 
& & \sum_{j \in \mathcal{L}} V_{\ell} (j, C_k) \\
& \text{subject to}
& & |\mathcal{L}| = L \\
& & & \mathcal{L} \cap \beta_u = \emptyset
\end{aligned}   
\end{equation}
The optimal solution is therefore obtained by selecting the $L$ products with the highest scores $V_{\ell}(j, C_k)$ that are not in the user's current basket.

\subsection{Evaluation Metrics}\label{evaluation metrics}

To assess the quality of the recommendation system, we employ the classical Precision metric, which measures the proportion of relevant items among the recommended ones. In addition, we use the more modern metrics Normalized Discounted Cumulative Gain (NDCG) and Normalized Discounted Cumulative Value (NDCV), which account not only for the relevance of the recommended items but also for their ranking positions. While NDCG focuses on recommendation relevance, NDCV extends this concept by incorporating the economic value (e.g., revenue or profit) associated with the recommended products, thus providing a more business-oriented evaluation of recommendation quality. 
\begin{itemize}
    \item \textbf{Precision} \citet{manning2008information}: It quantifies the proportion of correctly recommended items among the total number of recommended items $L$.  It reflects the system’s ability to generate accurate recommendations. Its formulation is:
    \[
    Precision@L = \frac{\text{Number of relevant items in the top-}L}{L}.
    \] 
     
   % \item  The \textbf{Recall} indicates the proportion of relevant items that are successfully recommended, relative to the total number of relevant items in the dataset. \textit{It reflects the system’s ability to recommend relevant content for the user}. It is defined as:
    % \[
    % \text{Recall@L} = \frac{\text{Number of relevant items in the top-}L}{\text{Number of relevant items}}.
    % \]     
    % \item The \textbf{Coverage} measures the proportion of unique items that the recommendation system is able to suggest to all users, relative to the total number of items available. \textit{It reflects the system’s ability to generate diverse recommendations across different users}. Its mathematical definition is:
    % \[
    % \text{Coverage} = \frac{\text{Number of items recommended across all users}}{\text{Total number of items}}.
    % \] 
    \item \textbf{Normalized Discounted Cumulative Gain} \citet{jarvelin2002}: It measures how well the system ranks the most relevant items purchased by the user in the highest positions. It is defined as:
    \begin{equation}
    NDCG@L = \frac{1}{|\mathcal{U}|} \sum_{u \in \mathcal{U}} \frac{DCG_{u}@L}{IDCG_{u}@L},
    \end{equation}
    where \( DCG_{u}@L \) is the \textbf{Discounted Cumulative Gain} at position \( L \) for user \( u \):
    \[
    DCG_{u}@L = \sum_{j=1}^{L} \frac{rel_{u,j}}{\log_2(j+1)},
    \]
    and $IDCG_{u}@L$ is the \textbf{Ideal Discounted Cumulative Gain}, obtained by sorting all the items relevant to the user \( u \) in descending order of relevance:
    \begin{equation}
    IDCG_{u}@L = \sum_{j=1}^{L} \frac{rel^{\ast}_{u,j}}{\log_2(j+1)},
    \end{equation}
    Here, $rel_{u,j}$ denotes the relevance score assigned to the $j$-th item for user $u$ in the predicted ranking, while $rel^{\ast}_{u,j}$ represents the relevance score in the ideal ranking. In this case, both scores are binary, indicating whether the product appears (1) or not (0) in the predicted or ideal ranking. The term $\frac{1}{\log_2(j+1)}$ is a \textbf{discount factor}, that reduces the contribution of items appearing in lower-ranked positions. Finally, $|\mathcal{U}|$ denotes the total number of users in the evaluation set. Observe that the $IDCG_{u}@L$ represents the maximum possible $DCG_{u}@L$, thus serving as the normalization factor for $NDCG$.

    \item \textbf{Normalized Discounted Cumulative Value} \citet{debiasio2023}: This metric extends the concept of NDCG by incorporating both user relevance and organizational value of items. It evaluate how well the system recommends items that are both relevant to users and valuable for the organization. The  \( NDCV@k \) at position \( k \) is defined as:
    \[
    NDCV@k = \frac{1}{| \mathcal{U} |} \sum_{u \in \mathcal{U}} \frac{DCV_{u}@L}{IDCV_u@k} 
    \]
    where $DCV_{u}@L$ is the \textbf{Discounted Cumulative Value} for user \(u\):
    \[
    DCV_{u}@L = \sum_{j=1}^L \frac{rel_{u,j} \cdot v_j}{\log_2(j+1)},
    \]
    and $IDCG_{u}@L$ denotes the \textbf{Ideal Discounted Cumulative Gain}, computed by sorting all relevant items user \(u\) in descending order of their value:
    
    \begin{equation}
    IDCG_{u}@L = \sum_{j=1}^{L} \frac{rel^{\ast}_{u,j}}{\log_2(j+1)}
    \end{equation} 
    
    In this formulation, $rel_{u,j}$ denotes the binary \textbf{relevance score} (1 if the item is relevant to user \(u\), 0 otherwise) assigned to the item $j$ for user $u$ in the ordered ranking of $L$ items, while $v_j$ represents the \textbf{item value}, corresponding to its sales revenue share.
\end{itemize}

\section{Simulation}\label{simulations}

In this section, we present a simulation study designed to evaluate the performance of our clustering methodology under various settings. First, we describe the experimental setup, including the data generation process and parameter configurations that characterize three distinct scenarios representing different customer behavior and conditions of interest. Next, we report the performance of the clustering algorithms on the simulated datasets using the four different similarity measures, \texttt{MADD}, \texttt{Jaccard}, \texttt{Cosine} and  \texttt{Euclidean}. To do it, we evaluate the accuracy in identifying the optimal number of clusters $k$ as well as we report the silhouette index to compare the obtained clusters with the true labels for the metrics. Finally, we analyze the recommendation performance across scenarios, considering different combinations of parameters $\theta$ and $\beta$, the similarity measures and the three proposed recommendation methods \texttt{Popularity}, \texttt{Revenue} and \texttt{ExpProfit}.

\subsection{Setting}
We consider a population composed of $K$ types of consumers, each exhibiting specific purchasing behavior driven by their product preferences and expenditure patterns. The preferences of each consumer type are characterized by three key parameters: 

\begin{itemize}
    \item {\bf Demand scale} $\alpha$, which controls the total expenditure on preferred products. {This parameter allows us to classify consumer types according to their total spending level. A consumer type with $\alpha = 0.1$ spends approximately 10\% as much as a consumer type with $\alpha = 1$.}
    \item {\bf Heterogeneity} {$\theta$, controls the fraction of preferred items that can co-occur within type $k$ (for $k = 1, \ldots, K$). The remaining $(1-\theta)$ proportion represents \textit{structural zeros}, i.e. items that are never purchased by the customer, even if they fall within their preferred range. When $\theta \rightarrow 1$, all customers of type $k$ have the same preferred basket, representing homogeneous preferences within each type.}   %which determines the proportion of preferred items that may appear within the same type $k$ (for $k = 1, \ldots, K$), The remaining $(1-\theta)\%$ proportion is the {\it structural zeros}, meaning they are never purchased by the customer, even if they fall within their preferred range.  
    \item {\bf Sparsity fraction} $\beta$, which specifies the proportion of preferred products that remain unpurchased (i.e., not observed in the data and used for evaluation). Higher values of $\beta$ lead to fewer observed purchases, making the recommendation task more challenging.  
\end{itemize}

In our setting, instead of fixing $\theta$ to a single value for all customers, we draw it uniformly from an interval $[\theta_{min}, \theta_{max}]$, which controls how broad or narrow the interests of the consumers are. This approach better reflects real markets, where not all consumers of the same type behave identically.
%While the sparsity parameter models these common {\it sampling zeros}, the heterogeneity parameter generates {\it structural zeros}, giving an additional sparsity level in the final user-item interaction matrix. 
On the other hand, sparsity is incorporated to emulate real-world user-item interaction matrices, where consumers often purchase only a subset of their preferred products. This behavior may arise from factors such as seasonality, product durability, stockpiling tendencies, or budget constraints. 

\subsection{Synthetic Data Generation}\label{data}

Consumer expenditures are generated in a multi-step process. First, we draw raw expenditure values from a uniform distribution $\mathcal{U}[0,M]$ with $M=10,000$, and then we scale them according to the demand scale $\alpha$ associated with each consumer type:

\begin{itemize}
    \item {\bf Type A:} Prefers the first 80\% of items (1 to $0.8p$), with low demand ($\alpha = 0.1$).
    \item {\bf Type B}: Prefers the last 10\% of items ($0.9p$ to $p$), with high demand ($\alpha =1$).
    \item {\bf Type C}: Prefers the second half of items ($0.6p$ to $0.95p$), with low-medium demand ($\alpha = 0.4$).
    \item {\bf Type D}: Prefers items in the middle range ($0.4p$ to $0.6p$), with low-medium demand ($\alpha = 0.4$).
\end{itemize} 

The total number of products is set to $p=1,500$. To capture intra-type heterogeneity, we simulate $n = 150$ individual consumers per type. For each consumer, only a fraction $\theta$ of the products within their preferred range is considered, with $\theta$ drawn from two possible intervals: $[0.50, 0.75]$ and $[0.85, 0.95]$. To model sparsity, the sparsity fraction $\beta$ is set to either $0.95$ or $0.99$. Finally, to account for occasional purchases of non-preferred products, we allow that 5\% of the purchases of each consumer to come from outside their preferred set. The expenditure for these purchases are reduced by a 20\% discount factor to reflect lower interest in these non-preferred products. 

Based on the consumer types defined previously, we analyze three scenarios depending on the number of clusters ($K$):

\begin{itemize}
    \item {\bf Scenario I ($K=2$):} Two highly distinct types, $A$ and $B$, representing the most disjoint purchasing behaviors.

    \item {\bf Scenario II ($K=3$):}  Three types $A$, $B$ and $C$, introducing intermediate heterogeneity. 

      \item {\bf Scenario III ($K=4$):} Four types $A$, $B$, $C$ and $D$,  capturing finer granularity in preferences. 
    
\end{itemize}

\begin{figure}
   \centering
   \includegraphics[width=\textwidth]{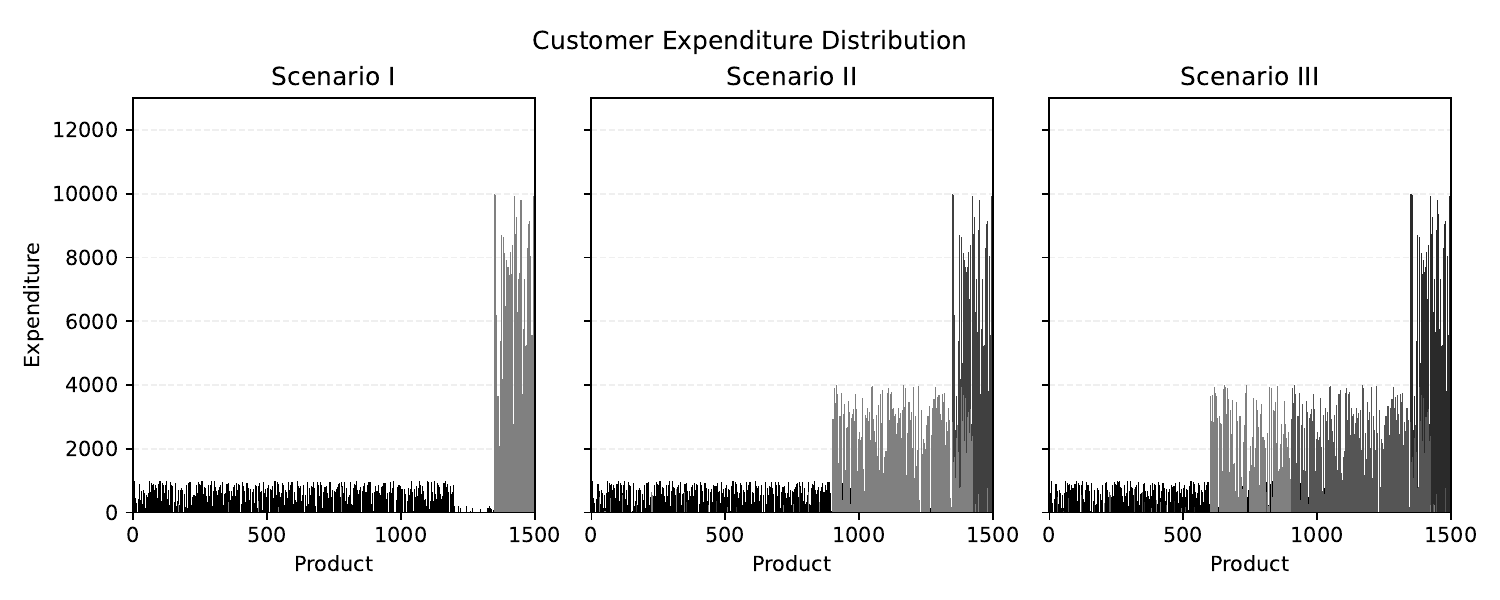}
   \caption{An example of the three scenarios and consumer types considered in the simulations.} \label{consumer_type}
\end{figure}

Train and test splits are defined according to the sparsity fraction $\beta$, which separates the \emph{products} into observed (training) and unobserved (testing) subsets for each consumer. Note that this means that all users are present in both training and test split. The resulting user-item interaction matrix has dimensions $(n \times K) \times p$. An illustrative example of the three scenarios and consumer types is shown in Figure \eqref{consumer_type}.

\subsection{Performance in clustering}\label{clustering}

Since the proposed recommendation system relies on clustering, we first evaluate the accuracy of our clustering method using the \texttt{MADD} distance applied to the user-item interaction matrix 
$\mathbf{S}$ of sales revenue share.  We compare its performance against clustering standard alternatives using \texttt{Euclidean} and \texttt{Cosine} distances on $\mathbf{S}$, or the \texttt{Jaccard} distance on the binary matrix $\mathbf{D}$. Once simulated consumer purchases under the three scenarios presented in Section \eqref{data}, we apply 
$K$-medoids clustering for each distance metric. Each experiment is repeated 50 times to ensure statistical reliability.

\begin{figure}
    \centering
  % Subfigura (A)
    \begin{subfigure}{0.88\textwidth}
        \centering
        \includegraphics[width=\textwidth]{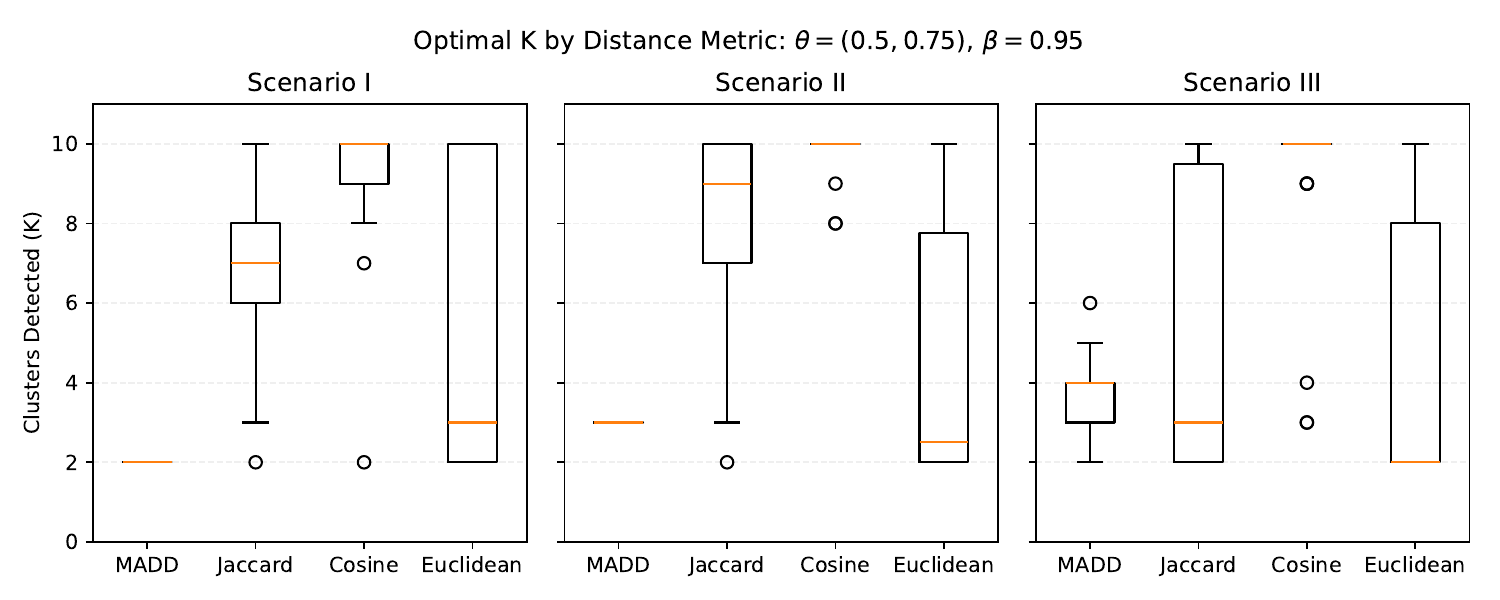}
        %%\label{subfig1}
    \end{subfigure}
        \vspace{-0.01cm}  
    \begin{subfigure}{0.88\textwidth}
        \centering        \includegraphics[width=\textwidth]{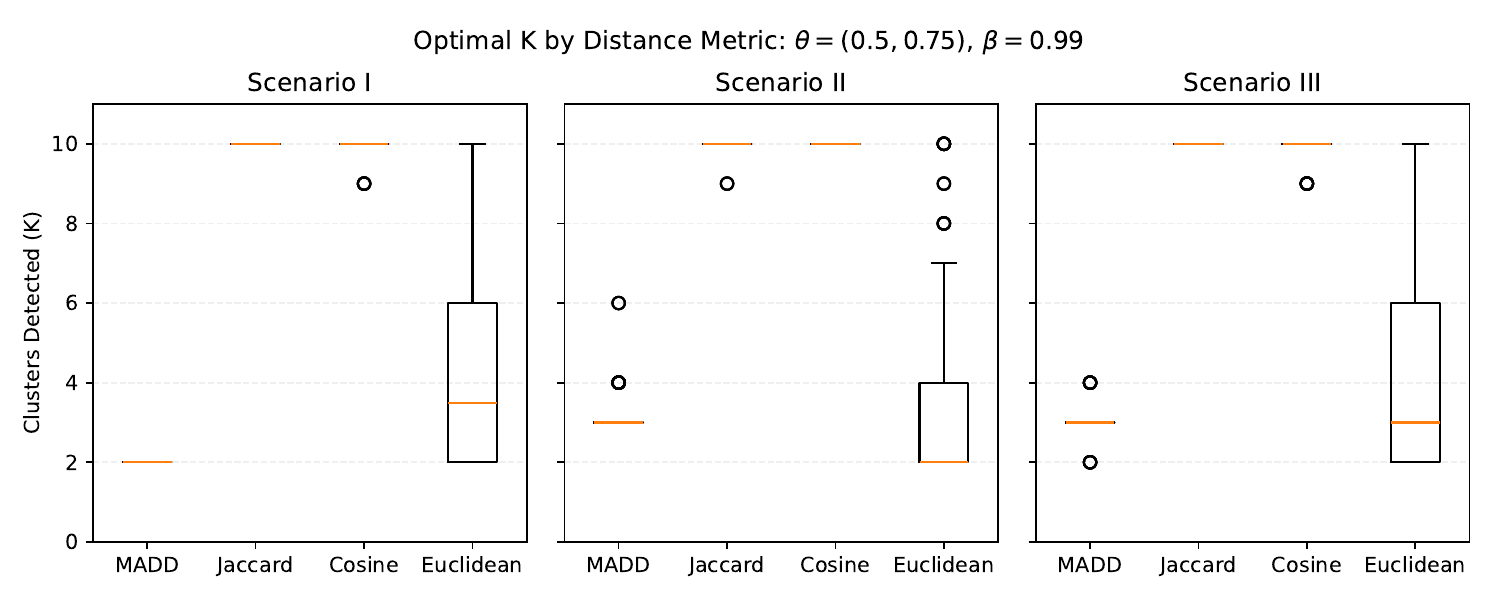}
        %\label{subfig2}
    \end{subfigure}
        \vspace{-0.01cm} 
    \begin{subfigure}{0.88\textwidth}
        \centering        \includegraphics[width=\textwidth]{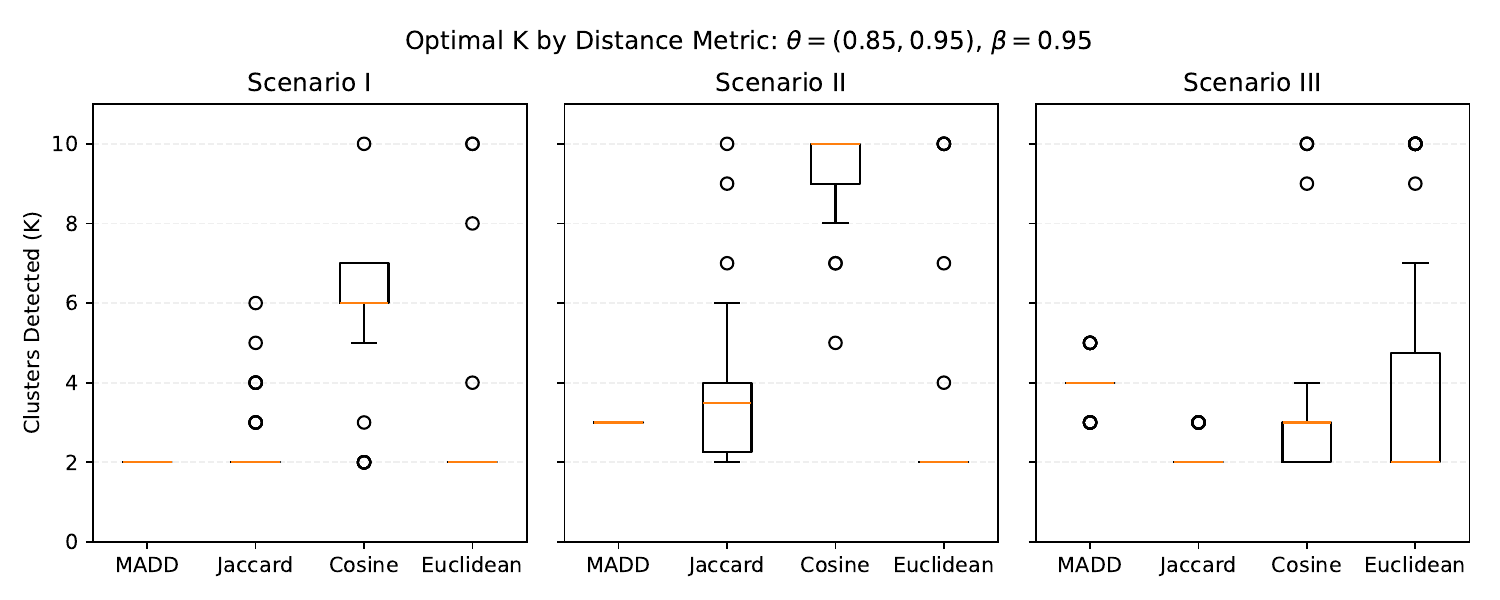}
        %\label{subfig3}
    \end{subfigure}
            \vspace{-0.01cm} 
        \begin{subfigure}{0.88\textwidth}
        \centering        \includegraphics[width=\textwidth]{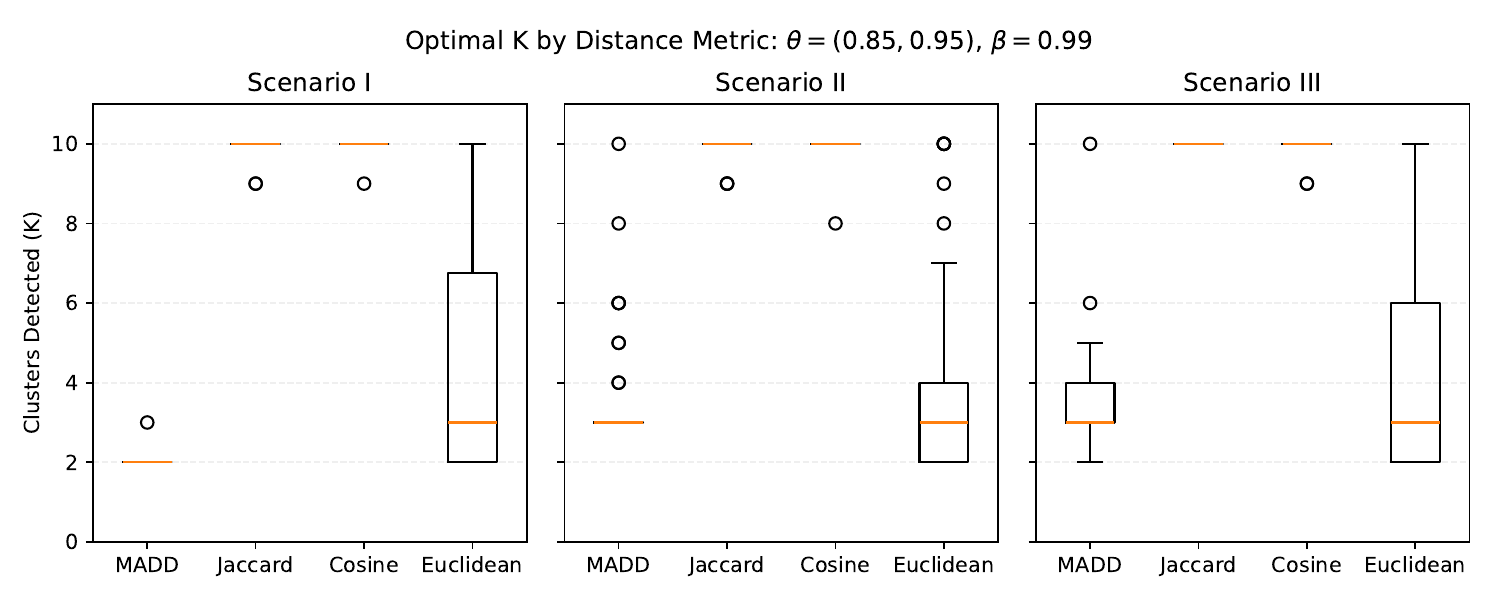}
        %\label{subfig3}
    \end{subfigure}
    \caption{Number of clusters selected in clustering for the different similarity measures, computed over 50 runs across different scenarios and parameters.}  
    \label{fig:clusterization}
\end{figure}

Figure \ref{fig:clusterization} presents box plots of the selected number of clusters obtained using alternative similarity measures within the $k$-medoids clustering algorithm, across different scenarios and parameter combinations. In all cases, the superior performance of the \texttt{MADD} similarity measure in correctly identifying the true number of clusters (2 for Scenario I, 3 for Scenario II, and 4 for Scenario III) is evident. More precisely, in most cases the \texttt{MADD} measure is almost exact in selecting $k$, although it becomes more erratic in the most complex Scenario III. Furthermore, it can be observed that the \texttt{Jaccard} and \texttt{Cosine} similarity measures deteriorate as sparsity increases, whereas \texttt{MADD} remains stable. In contrast, the \texttt{Euclidean} similarity produces the most variable results, showing greater instability in contexts of higher heterogeneity ($\theta \in [0.5, 0.75]$) and lower sparsity ($\beta = 0.95$). However, this distance performs better than \texttt{Jaccard} and \texttt{Cosine} with more remarkable good performance when sparsity is high (and, in some sense, when the effective dimensionality of the space is lower). Overall, the \texttt{Jaccard} and \texttt{Cosine} similarity measures tend to overestimate the number of clusters, a tendency that is reinforced under higher sparsity levels ($\beta = 0.99$).

\begin{figure}
    \centering
  % Subfigura (A)
    \begin{subfigure}{0.88\textwidth}
        \centering
        \includegraphics[width=\textwidth]{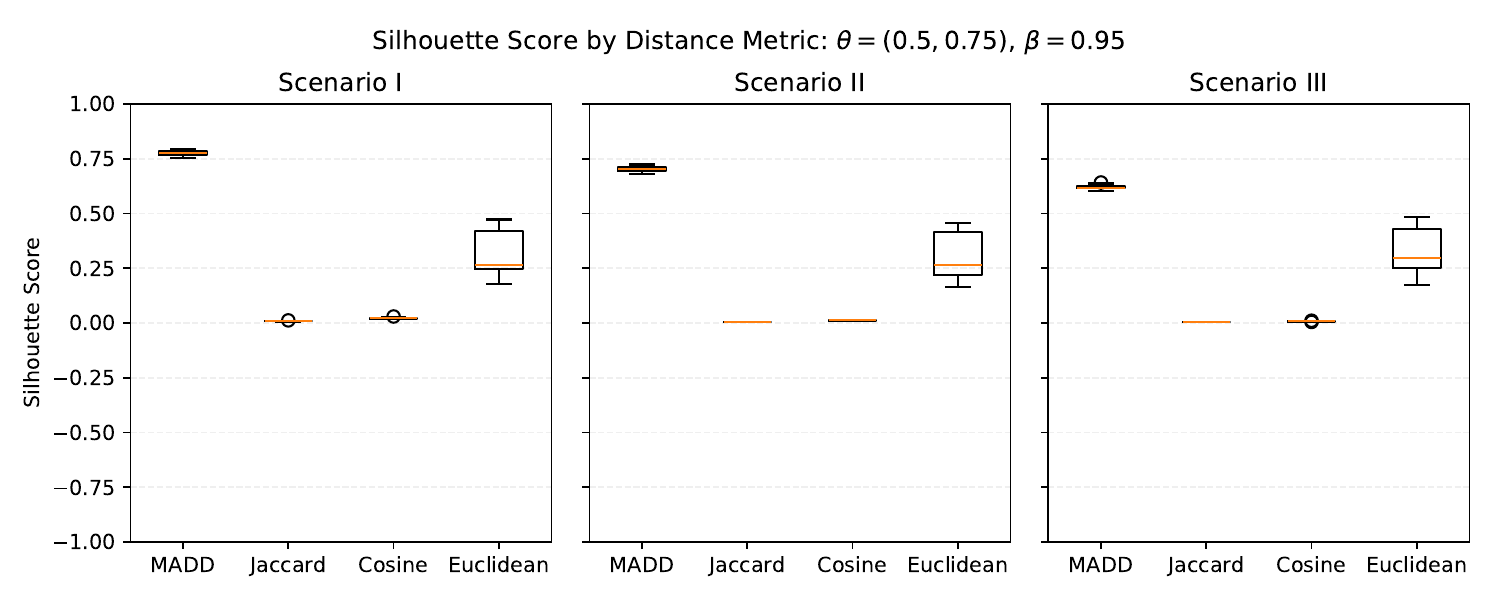}
        %\label{subfig1}
    \end{subfigure}
        \vspace{-0.01cm}  
    \begin{subfigure}{0.88\textwidth}
        \centering        \includegraphics[width=\textwidth]{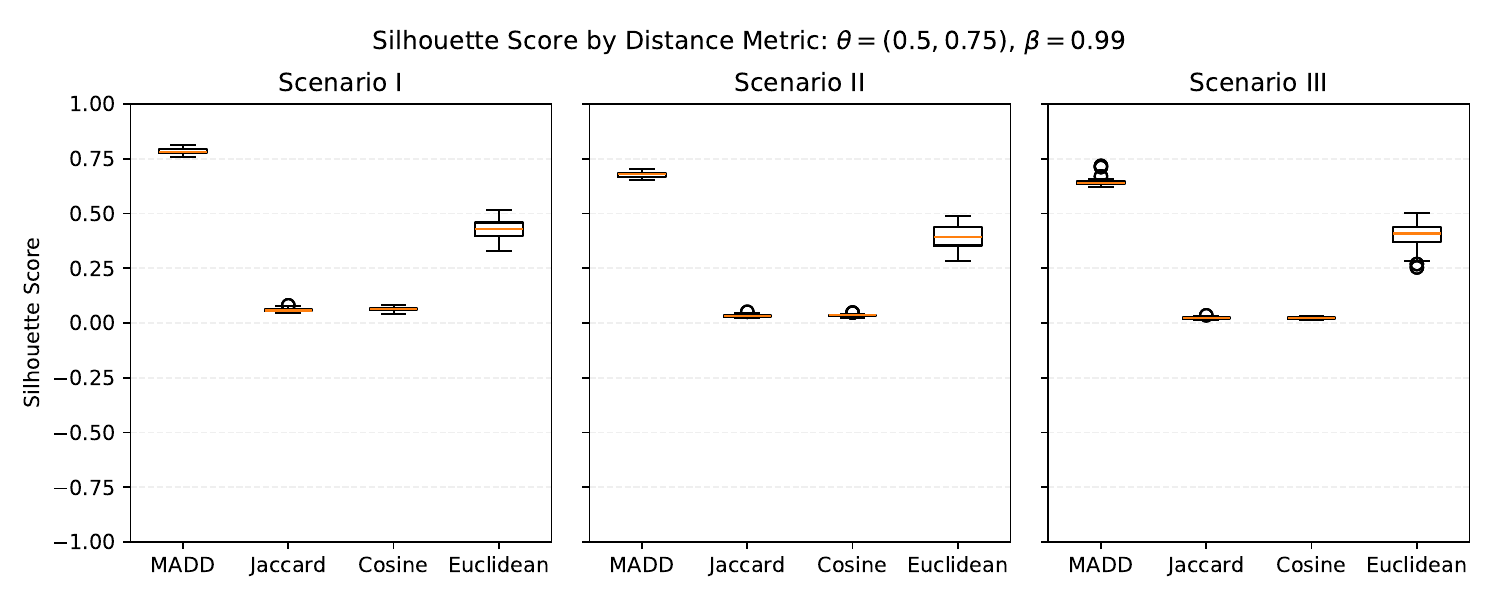}
        %\label{subfig2}
    \end{subfigure}
        \vspace{-0.01cm} 
    \begin{subfigure}{0.88\textwidth}
        \centering        \includegraphics[width=\textwidth]{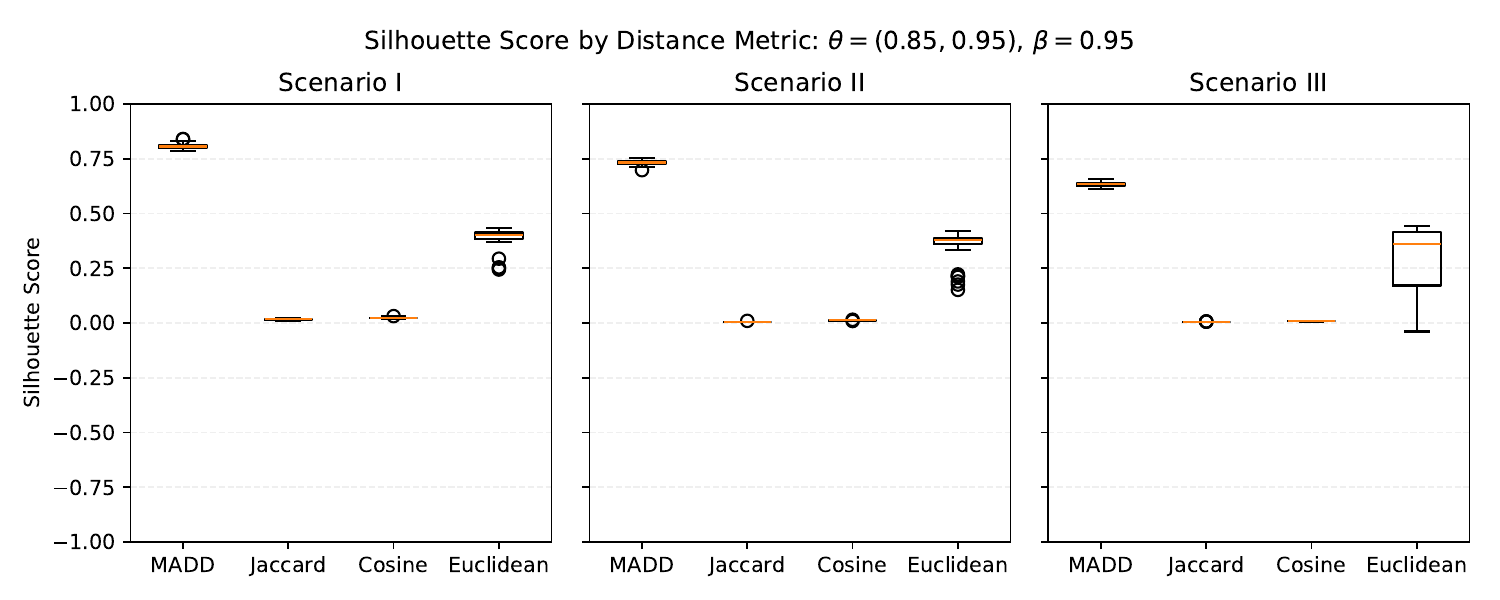}
        %\label{subfig3}
    \end{subfigure}
            \vspace{-0.01cm} 
        \begin{subfigure}{0.88\textwidth}
        \centering        \includegraphics[width=\textwidth]{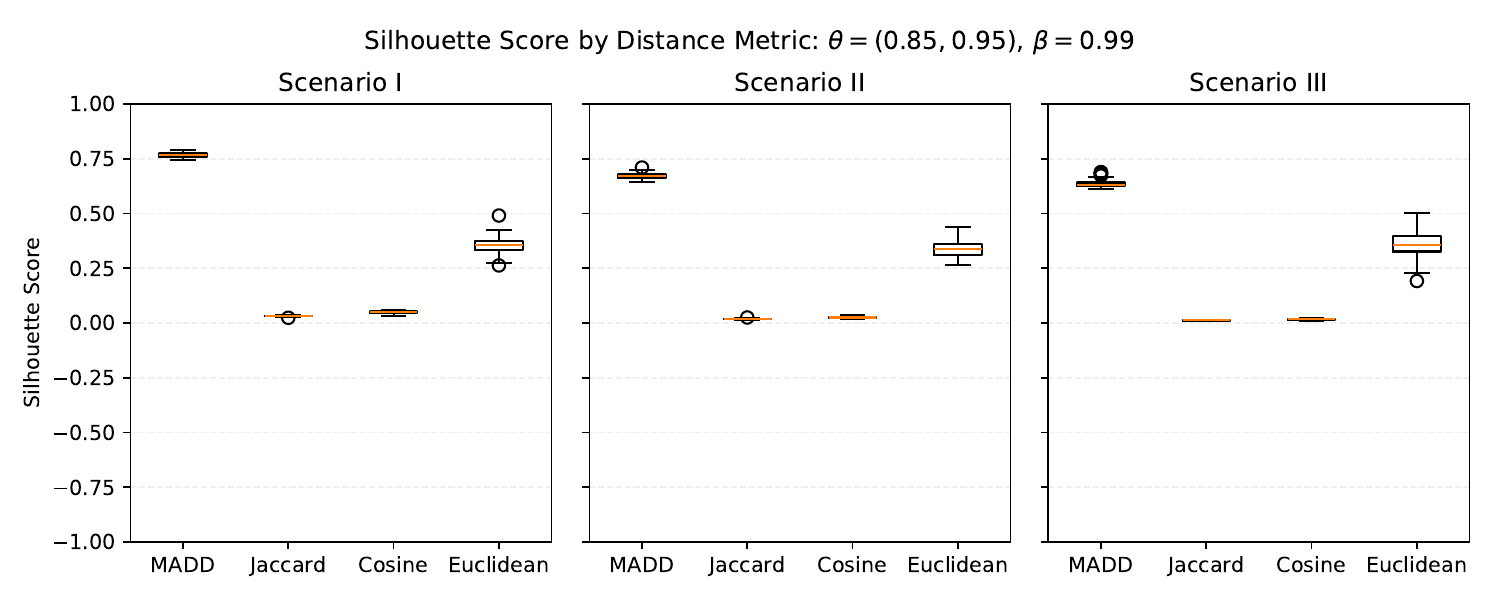}
        %\label{subfig3}
    \end{subfigure}
    \caption{Optimal Silhouette scores from clustering  for the different similarity measures, computed over 50 runs across different scenarios and parameters.}
    \label{fig:silhouette}
\end{figure}

The same pattern of behavior can be observed in the silhouette scores shown in the box plots of Figure \ref{fig:silhouette}, computed over the 50 runs. The \texttt{MADD} measure suggests a more appropriate clustering structure compared to the other three similarity measures, with a clearly better performance of the \texttt{Euclidean} distance over \texttt{Jaccard} and \texttt{Cosine}.

\subsection{Performance in recommendation}

In this section we evaluate accuracy of the three recommendation methods using the evaluation metrics Precision@L, NDCG@L, and NDCV@L. 

To perform this simulation, to each user in the evaluation set we recommend the top $10$ items that remain unpurchased (controlled then by the sparsity parameter $\beta$). To account for variability, we repeat the experiment of simulating the data and performing recommendation 50 times, then the mean and standard deviation across recommendation methods, distances and evaluation metrics are reported in Tables \ref{tab:scenario_III_5-75_95} to \ref{tab:scenario_III_85-95_99} for both selections of $\theta$ and $\beta$. 

In this section, we report the results for the more challenging Scenario, this is, for Scenario III, while results for Scenarios I and II are presented in the Appendix, although they have similar conclusions.

From Tables \ref{tab:scenario_III_5-75_95} and \ref{tab:scenario_III_5-75_99}, that is, when considering $\theta = (0.5, 0.75)$ fixed but allowing $\beta$ to change from 0.95 to 0.99, it can be observed that the \texttt{MADD} metric outperforms \texttt{Cosine} as sparsity increases. Only for the \texttt{Popularity} and \texttt{ExpProfit} recommendation methods with lower $\beta$ does \texttt{Cosine} outperform \texttt{MADD} in some metrics.

A clear winner behavior for the \texttt{MADD} distance with respect to $\beta$ can be observed in Tables \ref{tab:scenario_III_85-95_95} and \ref{tab:scenario_III_85-95_99}, where it achieves the best results in all cases.

In conclusion, just for a lower proportion of never purchased products, the \texttt{Cosine} distance outperforms \texttt{MADD}, for \texttt{Cosine}
\begin{table}[!ht]
\centering
\begin{tabular}{lcccc}
\hline
& & \multicolumn{3}{c}{Evaluation metrics} \\
\hline
Method & Distance & \texttt{Precision@10} & \texttt{NDCG@10} & \texttt{NDCV@10} \\ 
\hline
\multirow{4}{*}{\texttt{Popularity}}  
 & \texttt{MADD} & \textbf{0.483} (0.035) & 0.484 (0.036) & 0.674 (0.030) \\
 & \texttt{Jaccard} & 0.419 (0.057) & 0.424 (0.059) & 0.636 (0.065) \\
 & \texttt{Cosine} & 0.477 (0.031) & \textbf{0.485} (0.032) & \textbf{0.714} (0.041) \\
 & \texttt{Euclidean} & 0.312 (0.009) & 0.313 (0.010) & 0.573 (0.043) \\
\hline
\multirow{4}{*}{\texttt{Revenue}}  
 & \texttt{MADD} & \textbf{0.493} (0.038) & \textbf{0.496} (0.040) & \textbf{0.714} (0.032) \\
 & \texttt{Jaccard} & 0.374 (0.086) & 0.374 (0.087) & 0.594 (0.105) \\
 & \texttt{Cosine} & 0.461 (0.051) & 0.464 (0.052) & 0.711 (0.057) \\
 & \texttt{Euclidean} & 0.301 (0.019) & 0.303 (0.020) & 0.496 (0.018) \\
\hline
\multirow{4}{*}{\texttt{ExpProfit}}  
 & \texttt{MADD} & \textbf{0.493} (0.036) & \textbf{0.497} (0.037) & 0.706 (0.029) \\
 & \texttt{Jaccard} & 0.406 (0.076) & 0.407 (0.077) & 0.662 (0.061) \\
 & \texttt{Cosine} & 0.483 (0.042) & 0.487 (0.043) & \textbf{0.736} (0.028) \\
 & \texttt{Euclidean} & 0.322 (0.011) & 0.323 (0.012) & 0.518 (0.031) \\
\hline
\end{tabular}
\caption{Mean (and standard deviation in parentheses) of the Precision@10, NDCG@10 and NDCV@10 computed over 50 runs across recommendation methods and similarity measures for Scenario III with $\theta$ = (0.5, 0.75), $\beta$ = 0.95.}
\label{tab:scenario_III_5-75_95}
\end{table}

%----------------------------------------------
\begin{table}[!ht]
\centering
\begin{tabular}{lcccc}
\hline
& & \multicolumn{3}{c}{Evaluation metrics} \\
\hline
Method & Distance & \texttt{Precision@10} & \texttt{NDCG@10} & \texttt{NDCV@10} \\ 
\hline
\multirow{4}{*}{\texttt{Popularity}}  
 & \texttt{MADD} & \textbf{0.402} (0.017) & \textbf{0.403} (0.017) & \textbf{0.603} (0.032) \\
 & \texttt{Jaccard} & 0.313 (0.010) & 0.316 (0.012) & 0.573 (0.039) \\
 & \texttt{Cosine} & 0.313 (0.014) & 0.318 (0.014) & 0.586 (0.039) \\
 & \texttt{Euclidean} & 0.310 (0.010) & 0.311 (0.011) & 0.566 (0.046) \\
\hline
\multirow{4}{*}{\texttt{Revenue}}  
 & \texttt{MADD} & \textbf{0.402} (0.025) & \textbf{0.408} (0.027) & \textbf{0.669} (0.024) \\
 & \texttt{Jaccard} & 0.280 (0.025) & 0.285 (0.029) & 0.504 (0.036) \\
 & \texttt{Cosine} & 0.298 (0.024) & 0.302 (0.026) & 0.522 (0.032) \\
 & \texttt{Euclidean} & 0.281 (0.023) & 0.285 (0.026) & 0.487 (0.027) \\
\hline
\multirow{4}{*}{\texttt{ExpProfit}}  
 & \texttt{MADD} & \textbf{0.416} (0.022) & \textbf{0.419} (0.023) & \textbf{0.655} (0.020) \\
 & \texttt{Jaccard} & 0.301 (0.024) & 0.304 (0.027) & 0.564 (0.045) \\
 & \texttt{Cosine} & 0.310 (0.023) & 0.316 (0.027) & 0.550 (0.036) \\
 & \texttt{Euclidean} & 0.299 (0.018) & 0.303 (0.019) & 0.536 (0.042) \\
\hline
\end{tabular}
\caption{Mean (and standard deviation in parentheses) of the Precision@10, NDCG@10 and NDCV@10 computed over 50 runs across recommendation methods and similarity measures for Scenario III with $\theta$ = (0.5, 0.75), $\beta$ = 0.99.}
\label{tab:scenario_III_5-75_99}
\end{table}

%----------------------------------------------
\begin{table}[!ht]
\centering
\begin{tabular}{lcccc}
\hline
& & \multicolumn{3}{c}{Evaluation metrics} \\
\hline
Method & Distance & \texttt{Precision@10} & \texttt{NDCG@10} & \texttt{NDCV@10} \\ 
\hline
\multirow{4}{*}{\texttt{Popularity}}  
 & \texttt{MADD} & \textbf{0.771} (0.058) & \textbf{0.772} (0.058) & \textbf{0.831} (0.045) \\
 & \texttt{Jaccard} & 0.618 (0.060) & 0.621 (0.059) & 0.730 (0.040) \\
 & \texttt{Cosine} & 0.637 (0.068) & 0.641 (0.070) & 0.742 (0.057) \\
 & \texttt{Euclidean} & 0.460 (0.030) & 0.461 (0.031) & 0.669 (0.048) \\
\hline
\multirow{4}{*}{\texttt{Revenue}}  
 & \texttt{MADD} & \textbf{0.790} (0.051) & \textbf{0.793} (0.050) & \textbf{0.875} (0.037) \\
 & \texttt{Jaccard} & 0.527 (0.102) & 0.526 (0.097) & 0.683 (0.080) \\
 & \texttt{Cosine} & 0.536 (0.112) & 0.533 (0.111) & 0.700 (0.098) \\
 & \texttt{Euclidean} & 0.460 (0.035) & 0.459 (0.037) & 0.602 (0.038) \\
\hline
\multirow{4}{*}{\texttt{ExpProfit}}  
 & \texttt{MADD} & \textbf{0.785} (0.053) & \textbf{0.787} (0.053) & \textbf{0.865} (0.037) \\
 & \texttt{Jaccard} & 0.588 (0.088) & 0.586 (0.087) & 0.749 (0.051) \\
 & \texttt{Cosine} & 0.612 (0.092) & 0.614 (0.094) & 0.773 (0.054) \\
 & \texttt{Euclidean} & 0.490 (0.027) & 0.491 (0.026) & 0.619 (0.036) \\
\hline
\end{tabular}
\caption{Mean (and standard deviation in parentheses) of the Precision@10, NDCG@10 and NDCV@10 computed over 50 runs across recommendation methods and similarity measures for Scenario III with $\theta$ = (0.85, 0.95), $\beta$ = 0.95.}
\label{tab:scenario_III_85-95_95}
\end{table}

% ----------------------------------------
\begin{table}[!ht]
\centering
\begin{tabular}{lcccc}
\hline
& & \multicolumn{3}{c}{Evaluation metrics} \\
\hline
Method & Distance & \texttt{Precision@10} & \texttt{NDCG@10} & \texttt{NDCV@10} \\ 
\hline
\multirow{4}{*}{\texttt{Popularity}}  
 & \texttt{MADD} & \textbf{0.617} (0.029) & \textbf{0.619} (0.031) & \textbf{0.735} (0.034) \\
 & \texttt{Jaccard} & 0.465 (0.018) & 0.470 (0.022) & 0.700 (0.047) \\
 & \texttt{Cosine} & 0.469 (0.017) & 0.477 (0.018) & 0.709 (0.046) \\
 & \texttt{Euclidean} & 0.453 (0.014) & 0.457 (0.013) & 0.652 (0.052) \\
\hline
\multirow{4}{*}{\texttt{Revenue}}  
 & \texttt{MADD} & \textbf{0.619} (0.056) & \textbf{0.626} (0.060) & \textbf{0.799} (0.026) \\
 & \texttt{Jaccard} & 0.435 (0.030) & 0.440 (0.034) & 0.621 (0.032) \\
 & \texttt{Cosine} & 0.454 (0.032) & 0.459 (0.035) & 0.635 (0.034) \\
 & \texttt{Euclidean} & 0.417 (0.028) & 0.425 (0.027) & 0.591 (0.021) \\
\hline
\multirow{4}{*}{\texttt{ExpProfit}}  
 & \texttt{MADD} & \textbf{0.638} (0.043) & \textbf{0.643} (0.045) & \textbf{0.787} (0.024) \\
 & \texttt{Jaccard} & 0.460 (0.031) & 0.466 (0.034) & 0.664 (0.047) \\
 & \texttt{Cosine} & 0.472 (0.034) & 0.481 (0.036) & 0.661 (0.042) \\
 & \texttt{Euclidean} & 0.444 (0.025) & 0.451 (0.024) & 0.630 (0.044) \\
\hline
\end{tabular}
\caption{Mean (and standard deviation in parentheses) of the Precision@10, NDCG@10 and NDCV@10 computed over 50 runs across recommendation methods and similarity measures for Scenario III with $\theta$ = (0.85, 0.95), $\beta$ = 0.99.}
\label{tab:scenario_III_85-95_99}
\end{table}

\section{Real Data Application}

We also evaluate our method on real data using the UCI Online Retail II dataset \citet{online_retail_ii_502} %\footnote{D. Chen, \textit{Online Retail II} [Dataset]. UCI Machine Learning Repository. \url{https://doi.org/10.24432/C5CG6D}}. 
This dataset contains all transactions occurring for a UK-based and registered non-store online retailer between 01/12/2009 and 09/12/2011. For our experiments, we select all UK transactions between May and August 2011, as we observe that these months represent a stable sales period not influenced by major seasonal effects. 

We perform a random 50/50 split to separate the observed data (training set, used for clustering and recommendation) from the unobserved data (test set, used for evaluation). We include all customers who purchased at least 20 items. The split is performed by masking products that will result unobserved but expected to be recommended. The resulting user–item matrix contains $1,212$ customers and $2,912$ products, with $33,746$ transactions in the training set and $33,143$ in the test set. This results in an extremely sparse training matrix ($99.04\%$ sparsity). To account for variability, we repeat this process 50 times, thus evaluating our method under 50 different random splits.

Table \ref{online_retail} reports the average Precision@10, NDCG@10, and NDCV@10 values (and their standard deviations) computed over 50 runs for the Online Retail dataset. In this real-world scenario, the overall performance levels are lower than in the simulated experiments, which is expected given the inherent noise and sparsity of real purchase data.

For \texttt{Popularity} and \texttt{ExpProfit} recommendation methods, the \texttt{Jaccard} similarity achieves higher results overall. However, for the \texttt{Revenue} recommendation method, where the revenue share of items is relevant for recommendation, the \texttt{MADD} measure shows competitive and stable performance across all recommendation methods, ranking closely behind \texttt{Jaccard}, and outperforming \texttt{Cosine} and \texttt{Euclidean} similarities. 

On the other hand, the similarity measures \texttt{Cosine} and \texttt{Euclidean}, which are the most commonly employed measures in the recommender systems yield weaker results, likely due to their sensitivity to sparse and unbalanced purchase vectors. 

These findings suggest that while \texttt{MADD} remains a robust and general-purpose similarity measure, \texttt{Jaccard} can outperform it in some scenarios like in this real example.

Another significant observation in this example is the larger gap between \texttt{NDCV@10} and the other two metrics, \texttt{Precision@10} and \texttt{NDCG@10}, a pattern not observed in the simulation study. This suggests that taking into account the economic value of products tends to amplify the performance differences among methods, emphasizing the importance of evaluating both relevance and profitability in recommendation tasks.

\begin{table}[!ht]
\centering
\begin{tabular}{lcccc}
\hline
& & \multicolumn{3}{c}{Evaluation metrics} \\
\hline
Method & Distance & \texttt{Precision@10} & \texttt{NDCG@10} & \texttt{NDCV@10} \\ 
\hline
\multirow{4}{*}{\texttt{Popularity}}  & \texttt{MADD} & 0.135 (0.019) & 0.144 (0.019) & 0.383 (0.031) \\
 & \texttt{Jaccard} & \textbf{0.155} (0.023) & \textbf{0.166} (0.024) & \textbf{0.389} (0.038) \\
  & \texttt{Cosine} & 0.135 (0.020) & 0.147 (0.022) & 0.366 (0.036) \\
  & \texttt{Euclidean} & 0.137 (0.019) & 0.146 (0.019) & 0.381 (0.031) \\
\hline
\multirow{4}{*}{\texttt{Revenue}}  & \texttt{MADD} & \textbf{0.126} (0.019) & \textbf{0.138} (0.020) & \textbf{0.385} (0.033) \\
 & \texttt{Jaccard} & 0.114 (0.021) & 0.125 (0.023) & 0.345 (0.041) \\
  & \texttt{Cosine} & 0.098 (0.017) & 0.107 (0.018) & 0.302 (0.036) \\
  & \texttt{Euclidean} & 0.117 (0.020) & 0.128 (0.022) & 0.364 (0.038) \\
\hline
\multirow{4}{*}{\texttt{ExpProfit}}  & \texttt{MADD} & 0.131 (0.020) & 0.142 (0.020) & 0.388 (0.033) \\
 & \texttt{Jaccard} & \textbf{0.148} (0.023) & \textbf{0.159} (0.025) & \textbf{0.394} (0.041) \\
  & \texttt{Cosine} & 0.125 (0.019) & 0.137 (0.021) & 0.356 (0.036) \\
  & \texttt{Euclidean} & 0.130 (0.021) & 0.140 (0.022) & 0.382 (0.035) \\
\hline
\end{tabular}
\caption{Mean (and standard deviation in parentheses) of the Precision@10, NDCG@10 and NDCV@10 computed over 50 runs across recommendation methods and similarity measures for the Online Retail example.}
\label{online_retail}
\end{table}

\section{Conclusions} 
In this paper, we proposed and evaluated a robust similarity measure, the MADD distance, specifically designed to handle the challenges of data sparsity and high dimensionality in recommendation systems. As discussed, the performance of segmentation-based collaborative filtering methods strongly depends on the chosen similarity metric, making MADD a promising alternative to conventional measures such as Euclidean, cosine, and Jaccard. 

Our experiments on both simulated and real-world data demonstrate the value of this approach. For the recommendation step, we evaluated the standard top-$L$ algorithm against two value-maximizing alternatives. The results show that using the MADD distance with an expected profit criterion generally yields superior performance in terms of precision and NDCG. When the goal is to maximize NDCV, however, MADD paired with a pure revenue criterion is the most effective strategy. It is noteworthy that in our real-world case study, the Jaccard distance with expected profit also performed competitively, highlighting that discrete metrics can, in some contexts, outperform classical continuous-data metrics.

A key factor contributing to these results is our construction of the user-item matrix using revenue sales participation (each product's share of total revenue). This design offers critical advantages for value-aware clustering: it differentiates between high- and low-spending consumers; captures preference heterogeneity through budget allocation; is invariant to economy-wide price changes like inflation; and directly incorporates the revenue contribution of each product and segment, enabling value-based strategy design directly within the clustering step.

In summary, while the MADD distance generally proves superior for handling the sparse, high-dimensional data typical in RSs, the optimal combination of metric and recommendation criterion is context-dependent. Our work confirms that moving beyond classical similarity measures and incorporating nuanced, value-based data representations can significantly enhance the performance and business relevance of recommendation systems.

% ----------------------------------------
% \newpage 
% \bibliographystyle{apalike}
% \bibliography{biblio}
%\printbibliography
%[title={References}]

% ------------------------------------------

% ----------------------------------------
\newpage 
\appendix

\section{Additional results.}

Complementary tables for Scenarios I and II from Section \ref{simulations} of simulated data.

\

 \begin{table}[!ht]
\centering
\begin{tabular}{lcccc}
\hline
& & \multicolumn{3}{c}{Evaluation metrics} \\
\hline
Method & Distance & \texttt{Precision@10} & \texttt{NDCG@10} & \texttt{NDCV@10} \\ 
\hline
\multirow{4}{*}{\texttt{Popularity}}  & \texttt{MADD} & \textbf{0.581} (0.013) & \textbf{0.581} (0.013) & 0.763 (0.015) \\
 & \texttt{Jaccard} & 0.526 (0.024) & 0.532 (0.023) & 0.750 (0.019) \\
  & \texttt{Cosine} & 0.554 (0.032) & 0.558 (0.029) & \textbf{0.774} (0.018) \\
  & \texttt{Euclidean} & 0.326 (0.012) & 0.327 (0.013) & 0.570 (0.056) \\
\hline
\multirow{4}{*}{\texttt{Revenue}}  & \texttt{MADD} & 0.430 (0.039) & 0.416 (0.035) & 0.668 (0.038) \\
 & \texttt{Jaccard} & 0.498 (0.069) & 0.493 (0.073) & 0.722 (0.066) \\
  & \texttt{Cosine} & \textbf{0.531} (0.047) & \textbf{0.521} (0.056) & \textbf{0.749} (0.051) \\
  & \texttt{Euclidean} & 0.357 (0.008) & 0.357 (0.009) & 0.592 (0.014) \\
\hline
\multirow{4}{*}{\texttt{ExpProfit}}  & \texttt{MADD} & \textbf{0.570} (0.024) & \textbf{0.569} (0.027) & 0.775 (0.018) \\
 & \texttt{Jaccard} & 0.543 (0.020) & 0.541 (0.025) & 0.758 (0.023) \\
  & \texttt{Cosine} & 0.570 (0.019) & 0.569 (0.018) & \textbf{0.788} (0.017) \\
  & \texttt{Euclidean} & 0.357 (0.009) & 0.357 (0.010) & 0.586 (0.015) \\
\hline
\end{tabular}
\caption{Mean (and standard deviation in parentheses) of the Precision@10, NDCG@10 and NDCV@10 computed over 50 runs across recommendation methods and similarity measures  for Scenario I with $\theta$ = (0.5, 0.75), $\beta$ = 0.95.}
\label{tab:scenario_I_5-75_95}
\end{table}

\begin{table}[!ht]
\centering
\begin{tabular}{lcccc}
\hline
& & \multicolumn{3}{c}{Evaluation metrics} \\
\hline
Method & Distance & \texttt{Precision@10} & \texttt{NDCG@10} & \texttt{NDCV@10} \\ 
\hline
\multirow{4}{*}{\texttt{Popularity}}  & \texttt{MADD} & \textbf{0.495} (0.018) & \textbf{0.495} (0.018) & \textbf{0.643} (0.026) \\
 & \texttt{Jaccard} & 0.324 (0.008) & 0.329 (0.010) & 0.594 (0.054) \\
  & \texttt{Cosine} & 0.325 (0.009) & 0.332 (0.010) & 0.604 (0.057) \\
  & \texttt{Euclidean} & 0.322 (0.013) & 0.325 (0.012) & 0.584 (0.058) \\
\hline
\multirow{4}{*}{\texttt{Revenue}}  & \texttt{MADD} & \textbf{0.370} (0.009) & \textbf{0.370} (0.010) & \textbf{0.608} (0.019) \\
 & \texttt{Jaccard} & 0.308 (0.009) & 0.318 (0.009) & 0.582 (0.014) \\
  & \texttt{Cosine} & 0.314 (0.017) & 0.325 (0.016) & 0.594 (0.020) \\
  & \texttt{Euclidean} & 0.350 (0.020) & 0.351 (0.018) & 0.590 (0.018) \\
\hline
\multirow{4}{*}{\texttt{ExpProfit}}  & \texttt{MADD} & \textbf{0.421} (0.028) & \textbf{0.414} (0.028) & \textbf{0.676} (0.028) \\
 & \texttt{Jaccard} & 0.308 (0.009) & 0.318 (0.010) & 0.578 (0.017) \\
  & \texttt{Cosine} & 0.313 (0.017) & 0.324 (0.016) & 0.590 (0.019) \\
  & \texttt{Euclidean} & 0.350 (0.019) & 0.352 (0.017) & 0.584 (0.018) \\
\hline
\end{tabular}
\caption{Mean (and standard deviation in parentheses) of the Precision@10, NDCG@10 and NDCV@10 computed over 50 runs across recommendation methods and similarity measures  for Scenario I with $\theta$ = (0.5, 0.75), $\beta$ = 0.99.}
\label{tab:scenario_I_5-75_99}
\end{table}
\begin{table}[!ht]
\centering
\begin{tabular}{lcccc}
\hline
& & \multicolumn{3}{c}{Evaluation metrics} \\
\hline
Method & Distance & \texttt{Precision@10} & \texttt{NDCG@10} & \texttt{NDCV@10} \\ 
\hline
\multirow{4}{*}{\texttt{Popularity}}  & \texttt{MADD} & \textbf{0.870} (0.010) & \textbf{0.869} (0.010) & \textbf{0.902} (0.013) \\
 & \texttt{Jaccard} & 0.845 (0.022) & 0.845 (0.022) & 0.888 (0.023) \\
  & \texttt{Cosine} & 0.854 (0.014) & 0.855 (0.014) & 0.907 (0.016) \\
  & \texttt{Euclidean} & 0.463 (0.013) & 0.463 (0.013) & 0.669 (0.057) \\
\hline
\multirow{4}{*}{\texttt{Revenue}}  & \texttt{MADD} & 0.779 (0.061) & 0.751 (0.071) & \textbf{0.866} (0.048) \\
 & \texttt{Jaccard} & 0.625 (0.116) & 0.614 (0.107) & 0.775 (0.072) \\
  & \texttt{Cosine} & \textbf{0.780} (0.109) & \textbf{0.767} (0.116) & 0.874 (0.077) \\
  & \texttt{Euclidean} & 0.526 (0.008) & 0.526 (0.008) & 0.706 (0.013) \\
\hline
\multirow{4}{*}{\texttt{ExpProfit}}  & \texttt{MADD} & \textbf{0.870} (0.010) & \textbf{0.870} (0.010) & \textbf{0.927} (0.010) \\
 & \texttt{Jaccard} & 0.803 (0.075) & 0.782 (0.085) & 0.871 (0.055) \\
  & \texttt{Cosine} & 0.849 (0.030) & 0.845 (0.040) & 0.918 (0.030) \\
  & \texttt{Euclidean} & 0.526 (0.008) & 0.525 (0.008) & 0.700 (0.014) \\
\hline
\end{tabular}
\caption{Mean (and standard deviation in parentheses) of the Precision@10, NDCG@10 and NDCV@10 computed over 50 runs across recommendation methods and similarity measures  for Scenario I with $\theta$ = (0.85, 0.95), $\beta$ = 0.95.}
\label{tab:scenario_I_85-95_95}
\end{table}
\begin{table}[!ht]
\centering
\begin{tabular}{lcccc}
\hline
& & \multicolumn{3}{c}{Evaluation metrics} \\
\hline
Method & Distance & \texttt{Precision@10} & \texttt{NDCG@10} & \texttt{NDCV@10} \\ 
\hline
\multirow{4}{*}{\texttt{Popularity}}  & \texttt{MADD} & \textbf{0.768} (0.021) & \textbf{0.768} (0.022) & \textbf{0.794} (0.029) \\
 & \texttt{Jaccard} & 0.509 (0.018) & 0.520 (0.020) & 0.748 (0.048) \\
  & \texttt{Cosine} & 0.508 (0.014) & 0.522 (0.017) & 0.739 (0.056) \\
  & \texttt{Euclidean} & 0.475 (0.011) & 0.481 (0.011) & 0.684 (0.055) \\
\hline
\multirow{4}{*}{\texttt{Revenue}}  & \texttt{MADD} & \textbf{0.540} (0.019) & \textbf{0.539} (0.016) & 0.723 (0.016) \\
 & \texttt{Jaccard} & 0.460 (0.008) & 0.482 (0.007) & 0.722 (0.009) \\
  & \texttt{Cosine} & 0.479 (0.034) & 0.500 (0.032) & \textbf{0.734} (0.020) \\
  & \texttt{Euclidean} & 0.511 (0.021) & 0.516 (0.018) & 0.720 (0.010) \\
\hline
\multirow{4}{*}{\texttt{ExpProfit}}  & \texttt{MADD} & \textbf{0.664} (0.049) & \textbf{0.653} (0.052) & \textbf{0.812} (0.031) \\
 & \texttt{Jaccard} & 0.459 (0.008) & 0.482 (0.007) & 0.718 (0.010) \\
  & \texttt{Cosine} & 0.478 (0.033) & 0.500 (0.032) & 0.729 (0.019) \\
  & \texttt{Euclidean} & 0.511 (0.021) & 0.516 (0.018) & 0.711 (0.012) \\
\hline
\end{tabular}
\caption{Mean (and standard deviation in parentheses) of the Precision@10, NDCG@10 and NDCV@10 computed over 50 runs across recommendation methods and similarity measures  for Scenario I with $\theta$ = (0.85, 0.95), $\beta$ = 0.99.}
\label{tab:scenario_I_85-95_99}
\end{table}

%%%%%%%%%%%%%%%%%%%%%%%%%%%%%%%%%%%%%%%%%%%%%%%%%%%%%%%%%%
\begin{table}[!ht]
\centering
\begin{tabular}{lcccc}
\hline
& & \multicolumn{3}{c}{Evaluation metrics} \\
\hline
Method & Distance & \texttt{Precision@10} & \texttt{NDCG@10} & \texttt{NDCV@10} \\ 
\hline
\multirow{4}{*}{\texttt{Popularity}}  & \texttt{MADD} & \textbf{0.566} (0.012) & \textbf{0.567} (0.014) & \textbf{0.767} (0.022) \\
 & \texttt{Jaccard} & 0.489 (0.018) & 0.500 (0.019) & 0.734 (0.026) \\
  & \texttt{Cosine} & 0.480 (0.016) & 0.490 (0.016) & 0.731 (0.019) \\
  & \texttt{Euclidean} & 0.421 (0.011) & 0.421 (0.011) & 0.632 (0.044) \\
\hline
\multirow{4}{*}{\texttt{Revenue}}  & \texttt{MADD} & \textbf{0.569} (0.020) & \textbf{0.567} (0.026) & \textbf{0.797} (0.022) \\
 & \texttt{Jaccard} & 0.477 (0.036) & 0.478 (0.039) & 0.737 (0.039) \\
  & \texttt{Cosine} & 0.473 (0.026) & 0.477 (0.027) & 0.745 (0.023) \\
  & \texttt{Euclidean} & 0.398 (0.027) & 0.404 (0.027) & 0.655 (0.021) \\
\hline
\multirow{4}{*}{\texttt{ExpProfit}}  & \texttt{MADD} & \textbf{0.577} (0.013) & \textbf{0.580} (0.015) & \textbf{0.798} (0.014) \\
 & \texttt{Jaccard} & 0.507 (0.026) & 0.511 (0.027) & 0.754 (0.022) \\
  & \texttt{Cosine} & 0.497 (0.020) & 0.500 (0.021) & 0.755 (0.020) \\
  & \texttt{Euclidean} & 0.430 (0.018) & 0.432 (0.015) & 0.666 (0.013) \\
\hline
\end{tabular}
\caption{Mean (and standard deviation in parentheses) of the Precision@10, NDCG@10 and NDCV@10 computed over 50 runs across recommendation methods and similarity measures  for Scenario II with $\theta$ = (0.5, 0.75), $\beta$ = 0.95.}
\label{tab:scenario_II_5-75_95}
\end{table}

\begin{table}[!ht]
\centering
\begin{tabular}{lcccc}
\hline
& & \multicolumn{3}{c}{Evaluation metrics} \\
\hline
Method & Distance & \texttt{Precision@10} & \texttt{NDCG@10} & \texttt{NDCV@10} \\ 
\hline
\multirow{4}{*}{\texttt{Popularity}}  & \texttt{MADD} & \textbf{0.500} (0.015) & \textbf{0.502} (0.015) & \textbf{0.685} (0.029) \\
 & \texttt{Jaccard} & 0.379 (0.025) & 0.385 (0.026) & 0.624 (0.051) \\
  & \texttt{Cosine} & 0.374 (0.021) & 0.382 (0.021) & 0.642 (0.044) \\
  & \texttt{Euclidean} & 0.400 (0.022) & 0.404 (0.021) & 0.641 (0.052) \\
\hline
\multirow{4}{*}{\texttt{Revenue}}  & \texttt{MADD} & \textbf{0.453} (0.029) & \textbf{0.455} (0.029) & \textbf{0.733} (0.027) \\
 & \texttt{Jaccard} & 0.335 (0.029) & 0.342 (0.032) & 0.622 (0.036) \\
  & \texttt{Cosine} & 0.348 (0.027) & 0.357 (0.030) & 0.639 (0.036) \\
  & \texttt{Euclidean} & 0.371 (0.033) & 0.379 (0.035) & 0.643 (0.036) \\
\hline
\multirow{4}{*}{\texttt{ExpProfit}}  & \texttt{MADD} & \textbf{0.497} (0.017) & \textbf{0.498} (0.021) & \textbf{0.731} (0.021) \\
 & \texttt{Jaccard} & 0.365 (0.027) & 0.372 (0.032) & 0.655 (0.040) \\
  & \texttt{Cosine} & 0.368 (0.025) & 0.374 (0.028) & 0.656 (0.032) \\
  & \texttt{Euclidean} & 0.394 (0.029) & 0.400 (0.029) & 0.660 (0.027) \\
\hline
\end{tabular}
\caption{Mean (and standard deviation in parentheses) of the Precision@10, NDCG@10 and NDCV@10 computed over 50 runs across recommendation methods and similarity measures  for Scenario II with $\theta$ = (0.5, 0.75), $\beta$ = 0.99.}
\label{tab:scenario_II_5-75_99}
\end{table}

\begin{table}[!ht]
\centering
\begin{tabular}{lcccc}
\hline
& & \multicolumn{3}{c}{Evaluation metrics} \\
\hline
Method & Distance & \texttt{Precision@10} & \texttt{NDCG@10} & \texttt{NDCV@10} \\ 
\hline
\multirow{4}{*}{\texttt{Popularity}}  & \texttt{MADD} & \textbf{0.767} (0.021) & \textbf{0.771} (0.022) & \textbf{0.830} (0.028) \\
 & \texttt{Jaccard} & 0.563 (0.034) & 0.573 (0.034) & 0.761 (0.044) \\
  & \texttt{Cosine} & 0.559 (0.026) & 0.573 (0.027) & 0.765 (0.048) \\
  & \texttt{Euclidean} & 0.597 (0.022) & 0.602 (0.021) & 0.734 (0.054) \\
\hline
\multirow{4}{*}{\texttt{Revenue}}  & \texttt{MADD} & \textbf{0.712} (0.039) & \textbf{0.712} (0.047) & \textbf{0.873} (0.036) \\
 & \texttt{Jaccard} & 0.506 (0.043) & 0.518 (0.049) & 0.752 (0.043) \\
  & \texttt{Cosine} & 0.527 (0.039) & 0.542 (0.040) & 0.773 (0.032) \\
  & \texttt{Euclidean} & 0.554 (0.050) & 0.565 (0.050) & 0.770 (0.037) \\
\hline
\multirow{4}{*}{\texttt{ExpProfit}}  & \texttt{MADD} & \textbf{0.767} (0.028) & \textbf{0.772} (0.031) & \textbf{0.877} (0.024) \\
 & \texttt{Jaccard} & 0.546 (0.039) & 0.556 (0.045) & 0.780 (0.034) \\
  & \texttt{Cosine} & 0.560 (0.035) & 0.575 (0.036) & 0.790 (0.028) \\
  & \texttt{Euclidean} & 0.589 (0.040) & 0.598 (0.039) & 0.781 (0.024) \\
\hline
\end{tabular}
\caption{Mean (and standard deviation in parentheses) of the Precision@10, NDCG@10 and NDCV@10 computed over 50 runs across recommendation methods and similarity measures  for Scenario II with $\theta$ = (0.85, 0.95), $\beta$ = 0.95.}
\label{tab:scenario_II_85-95_95}
\end{table}

\begin{table}[!ht]
\centering
\begin{tabular}{lcccc}
\hline
& & \multicolumn{3}{c}{Evaluation metrics} \\
\hline
Method & Distance & \texttt{Precision@10} & \texttt{NDCG@10} & \texttt{NDCV@10} \\ 
\hline
\multirow{4}{*}{\texttt{Popularity}}  & \texttt{MADD} & \textbf{0.767} (0.021) & \textbf{0.771} (0.022) & \textbf{0.830} (0.028) \\
 & \texttt{Jaccard} & 0.563 (0.034) & 0.573 (0.034) & 0.761 (0.044) \\
  & \texttt{Cosine} & 0.559 (0.026) & 0.573 (0.027) & 0.765 (0.048) \\
  & \texttt{Euclidean} & 0.597 (0.022) & 0.602 (0.021) & 0.734 (0.054) \\
\hline
\multirow{4}{*}{\texttt{Revenue}}  & \texttt{MADD} & \textbf{0.712} (0.039) & \textbf{0.712} (0.047) & \textbf{0.873} (0.036) \\
 & \texttt{Jaccard} & 0.506 (0.043) & 0.518 (0.049) & 0.752 (0.043) \\
  & \texttt{Cosine} & 0.527 (0.039) & 0.542 (0.040) & 0.773 (0.032) \\
  & \texttt{Euclidean} & 0.554 (0.050) & 0.565 (0.050) & 0.770 (0.037) \\
\hline
\multirow{4}{*}{\texttt{ExpProfit}}  & \texttt{MADD} & \textbf{0.767} (0.028) & \textbf{0.772} (0.031) & \textbf{0.877} (0.024) \\
 & \texttt{Jaccard} & 0.546 (0.039) & 0.556 (0.045) & 0.780 (0.034) \\
  & \texttt{Cosine} & 0.560 (0.035) & 0.575 (0.036) & 0.790 (0.028) \\
  & \texttt{Euclidean} & 0.589 (0.040) & 0.598 (0.039) & 0.781 (0.024) \\
\hline
\end{tabular}
\caption{Mean (and standard deviation in parentheses) of the Precision@10, NDCG@10 and NDCV@10 computed over 50 runs across recommendation methods and similarity measures  for Scenario II with $\theta$ = (0.85, 0.95), $\beta$ = 0.99.}
\label{tab:scenario_II_85-95_99}
\end{table}
\end{document}